\documentclass[sigconf, nonacm, timestamp]{acmart}

\AtBeginDocument{%
  \providecommand\BibTeX{{%
      \normalfont B\kern-0.5em{\scshape i\kern-0.25em b}\kern-0.8em\TeX}}}

\usepackage{graphicx}
\usepackage{textcomp}
\usepackage{multirow}
\usepackage{color}
\usepackage[ruled, vlined]{algorithm2e}
\usepackage{epstopdf}
\usepackage{comment}
\usepackage{multirow}
\usepackage{url}

\usepackage[draft]{minted}

\newcommand{\one}{({\em i})\/}
\newcommand{\two}{({\em ii})\/}
\newcommand{\three}{({\em iii})\/}
\newcommand{\four}{({\em iv})\/}

\newcommand\s[1]{(\S\ref{#1})\xspace}

\begin{document}

\title{
  Data Management for Building Information Modelling \\
  in a Real-Time Adaptive City Platform}


\author{Justas Brazauskas}
\email{jb2328@cam.ac.uk}
\affiliation{%
  \institution{University of Cambridge}
  \city{Cambridge}
  \country{United Kingdom}
}
\author{Rohit Verma}
\email{rv355@cam.ac.uk}
\affiliation{%
  \institution{University of Cambridge}
  \city{Cambridge}
  \country{United Kingdom}
}
\author{Vadim Safronov}
\email{vs451@cam.ac.uk}
\affiliation{%
  \institution{University of Cambridge}
  \city{Cambridge}
  \country{United Kingdom}
}
\author{Matthew Danish}
\email{mrd45@cam.ac.uk}
\affiliation{%
  \institution{University of Cambridge}
  \city{Cambridge}
  \country{United Kingdom}
}
\author{Jorge Merino}
\email{jm2210@cam.ac.uk}
\affiliation{%
  \institution{University of Cambridge}
  \city{Cambridge}
  \country{United Kingdom}
}
\author{Xiang Xie}
\email{xx809@cam.ac.uk}
\affiliation{%
  \institution{University of Cambridge}
  \city{Cambridge}
  \country{United Kingdom}
}
\author{Ian Lewis}
\email{ijl20@cam.ac.uk}
\affiliation{%
  \institution{University of Cambridge}
  \city{Cambridge}
  \country{United Kingdom}
}
\author{Richard Mortier}
\email{rmm1002@cam.ac.uk}
\affiliation{%
  \institution{University of Cambridge}
  \city{Cambridge}
  \country{United Kingdom}
}


\begin{abstract}
  Legacy Building Information Modelling (BIM) systems are not designed to process the high-volume, high-velocity data emitted by in-building Internet-of-Things (IoT) sensors. Historical lack of consideration for the real-time nature of such data means that outputs from such BIM systems typically lack the timeliness necessary for enacting decisions as a result of patterns emerging in the sensor data. Similarly, as sensors are increasingly deployed in buildings, antiquated Building Management Systems (BMSs) struggle to maintain functionality as interoperability challenges increase. In combination these motivate the need to fill an important gap in smart buildings research, to enable faster adoption of these technologies, by combining BIM, BMS and sensor data. This paper describes the data architecture of the Adaptive City Platform, designed to address these combined requirements by enabling integrated BIM and real-time sensor data analysis across both time and space.
\end{abstract}



\maketitle

\section{Introduction}

Smart infrastructure projects, e.g.,~building automation, are increasingly dependent on BIM and BMS for metadata and data collection, analysis and activation components. BMS' often have the ability to monitor lighting, heating, ventilation and air conditioning (HVAC) as well as electricity consumption. Typically proprietary software, most BMS vendors provide closed system products with industrial interfaces for activation, control, and basic data visualisation without the additional contextual data from BIM. Simultaneously, considerable effort is being expended on the deployment of IoT devices to increase sensor density, but unfortunately the industry remains highly fragmented~\cite{koh2018scrabble,becerik2012application}.

Additionally, legacy BMS' and much current research focuses on in-building sensor data collection, storage and presentation platforms, rarely emphasising the challenges and benefits of being able to analyse and respond to data in real-time~\cite{tang2019review, dave2018framework,chevallierreference}. BMS' have historically dealt with low-volume low-velocity data and metadata, so the adoption of IoT devices poses substantial network and system challenges in dealing with real-time data analysis, event recognition, prediction and action planning~\cite{ramprasad2018leveraging}.

In this paper we focus on the real-time aspects of spatio-temporal data available from IoT sensing. We define a real-time platform as an asynchronous system capable of processing high-volume, high-heterogeneity data with minimal latency to collect, analyse, predict and adapt to changes in a timely manner.

A real-time data architecture is only part of the puzzle though: despite the increasing deployment of IoT devices, there are still no canonical means to join BIM and deployed sensors in a single unified system. While numerous attempts exist in the form of creating ontologies (e.g.,~BRICK)~\cite{balaji2016brick,balaji2018brick} to unify static metadata management for use by building automation systems, industry recognition for metadata standards is limited~\cite{koh2018scrabble,bhattacharya2015short, becerik2012application}. Also, as a result augmenting BIM with IoT devices, building and facility management software must be adapted~\cite{jourdan2019towards}. Highly siloed BMS software must become able to handle an increased amount of contextual building data in a timely manner to comply with the use of edge computing to for accident and emergency management~\cite{muralidharaair} and smart home initiatives resulting in the creation of safer and more resilient smart spaces~\cite{gao2019real,tapia2004activity}. New approaches that combine BIM, BMS and sensor data are thus needed.

To meet these important challenges, we propose the Adaptive City Platform (ACP), a system for collecting, processing and visualising building information and sensor data in real-time. Our contributions are: \one~the design of the ACP, a real-time software architecture combining BIM, BMS and IoT data into a unified low-latency, high-throughput building monitoring system; \two~description of our prototype implementation of the ACP, deployed in our department building; and \three~initial evaluation of our platform, in the form of a set of data management practices and detailed examples used to create a functioning front-end visualisation prototype combining heterogeneous data from BIM, BMS and IoT systems.

We first discuss the background and trends in BIM, BMS and IoT integration research in~\s{background} and describe the importance of a real-time platform in~\s{realtime}, followed by our original work presented in~\s{design}. We then discuss ACP's advantages as well as challenges it poses in areas such as privacy and sensor-derived constraints in~\s{discussion}, before concluding~\s{concls}.

\section{Related Work}\label{background}

Building Information Modelling (BIM) first appeared in the 1970s~\cite{eastman1974outline}, and is the design of buildings and infrastructure projects so as to enable different AEC (Architecture, Engineering, Construction) stakeholders to collaborate on a project~\cite{azhar2011building}. BIM models typically comprise a database of complete structural schematics encompassing 2D plans, 3D models, as well as data about building internals, e.g.,~the materials and parts used in construction. They have been found to provide many benefits, including more efficient collaboration and coordination in all phases of building construction, improved planning, reduced construction costs, and a lower carbon footprint~\cite{bryde2013project, azhar2008building, antonopoulou2017bim}. As a result, BIM use has increased five-fold in the last decade~\cite{nbs2019national}; the most popular BIM software today, used by over 46\% of architects and engineers in the industry, is Revit followed by Archicad~\cite{nbs2019national}. However, despite this rise in popularity and BIM's numerous advantages, its definition remains debated~\cite{doan2019bim}.

\begin{figure}
  \centering
  \includegraphics[width=.8\linewidth]{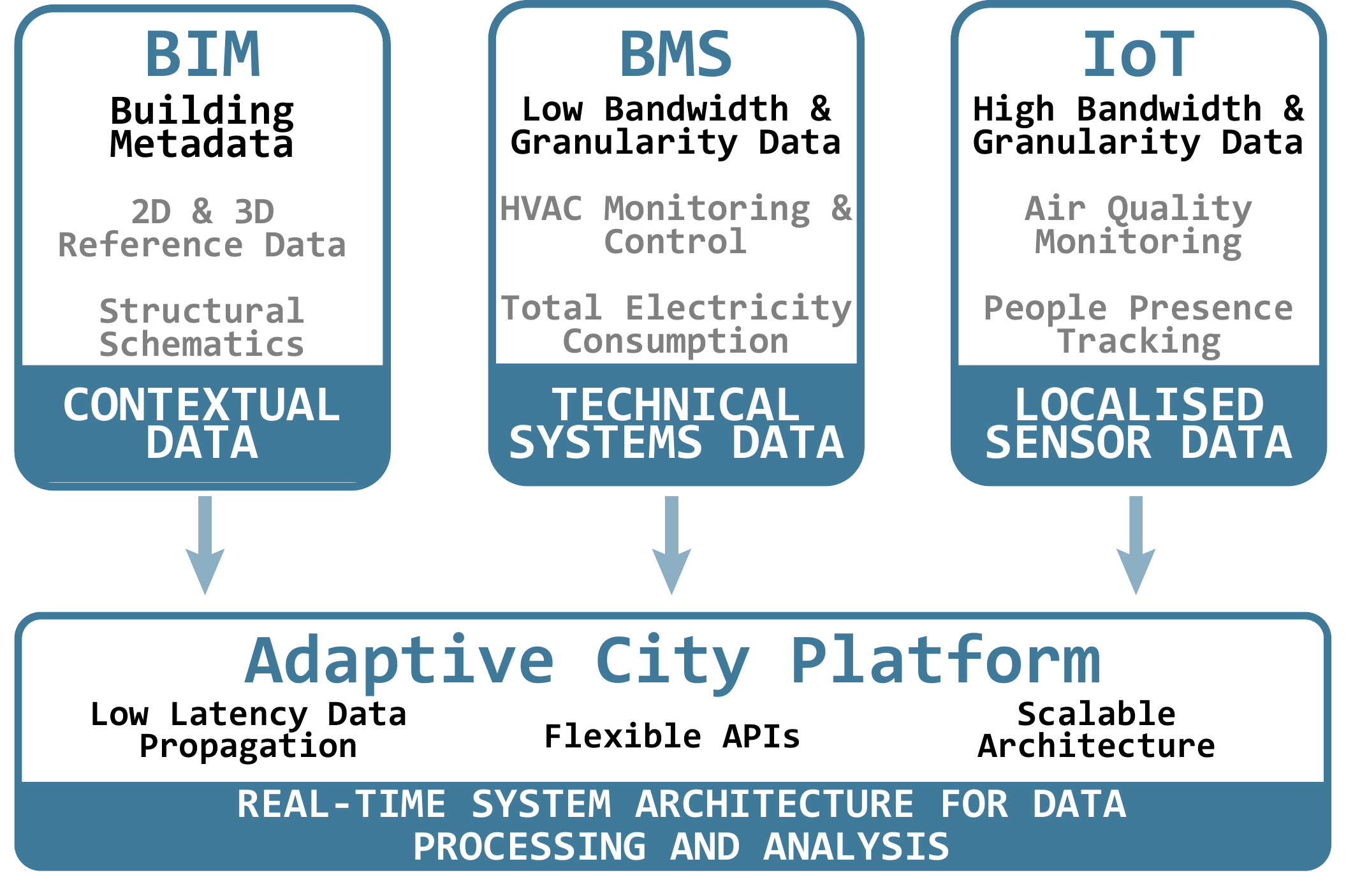}
  \caption{\label{bim_bms_iot}BIM-IoT fusion envisioned as managed in a single real-time platform combining the contextual, building-systems and localised data provided by BIM, BMS and IoT.}
\end{figure}

As well as becoming an integral part of complete building lifecycle management including planning, construction and maintenance phases, BIM is expected to play a key role in the deployment of IoT devices in smart buildings~\cite{cantarero2018common}. As the number of connected devices increases, around 11 billion IoT devices are expected to be installed in buildings worldwide~\cite{memoori2018}. However, deployment of these sensors presents challenges to do with the timeliness and contextualisation of the data they produce. To create smart, automated infrastructure, these IoT devices must interface with BIM and BMS platforms to provide both visualisation and actuation capabilities within the temporal and spatial building context. This three-way BIM-BMS-IoT (abbreviated as BIM-IoT) relation forms closed loop systems that we term BIM-IoT fusion (Figure~\ref{bim_bms_iot}).

Today, building systems monitoring and actuation is managed by highly encapsulated, siloed BMS systems~\cite{koh2018scrabble, becerik2012application}. Usually employed to monitor building-scale mechanical and electrical systems (e.g.,~HVAC, electricity consumption), the majority of BMS' are unfit for use with IoT as they typically lack the contextual information provided by the BIM software as well as the capacity to access and process low-latency, high-bandwidth IoT data~\cite{manic2016building, ramprasad2018leveraging}. Though there are numerous benefits for further integration of these technologies, BIM-IoT fusion remains in its infancy.

Successful BIM and IoT device use for energy, health, safety and building management, and construction monitoring have been demonstrated~\cite{teizer2017internet, park2017framework}. Furthermore, combined with BMS, BIM-IoT fusion allows for a new way to manage, control and ultimately automate building systems to achieve lower electricity consumption and reduce the CO\textsubscript{2} footprint~\cite{mousa2016utilizing}. While much smart infrastructure research focuses on overcoming the standardization problems and crafting ontologies for static reference data, BIM-IoT incorporation is becoming an important topic in its own right~\cite{tang2019review}.

While real-time analysis of data processing is often overlooked, several papers have proposed interlinking and contextualising sensor data using BIM tools~\cite{tang2019review}. Nevertheless, most attempts remain relatively small scale and often lack good (non-blocking, asynchronous) support for real-time data. A common pitfall with data storage-centric models that rely on repeated querying is that sensor data is collected and archived while sensor metadata is made an extension of the BIM model. This results in smart building platforms that are dependent on static BIM software with no real capacity to provide a low-latency data flow, as the examples below show.

Penna et al~\cite{penna2019sensors} describes a system where sensor readings are linked with a BIM database using Revit and Dynamo (a visual scripting language for Revit). The proposed system integrated LoRaWAN-powered environmental sensor data such as temperature, CO\textsubscript{2}, humidity with people's presence, storing it in a SQL database. The data was then loaded to Revit using a Dynamo script to visualise readings. However, the system lacked a real time component, resulting in transmission latencies of up to 15\,minutes.

Rasmussen et al~\cite{holten2018integrating, rasmussen2017web} presents an integrated web system using their proposed Linked Building Data ontology with sensor and actuator readings. Although their demo featured a well described web interface and detailed API reference, it lacked a real-time component as a result of their storage-centric model, as well as also relying on Revit. Similarly, Chevallier et al~\cite{chevallierreference} propose a reference architecture for Digital Twin buildings, using BIM and IoT technologies. Focused on ontology definitions, the presented system architecture and examples also feature a database-centric model reliant on queries.

Dave et al~\cite{dave2018framework} proposed a platform that integrates IoT data and real-time web visualisation environment, \emph{Otaniemi3D}, that showed the University of Aalto’s campus on several scales. They describe the design criteria, system architecture, and workflow, along with examples of how their system would operate in a real-world environment. However, this also assumed a ``collect, store, query'' model for sensor data which fundamentally limits its real-time capability. Additionally, the static nature of their XML-based data storage makes it unsuitable for spatio-temporal data.

More widely, Tang et al~\cite{tang2019review} provides an overview of research covering both BIM and IoT, and highlighting several major challenges in research regarding smart buildings. Key overlooked aspects identified include practical implementability -- many described systems were conceptual rather than usable. Moreover, prototypes were often tested under lab conditions, limiting the assessment of the ability of those systems to withstand the challenges of real-world deployments. Finally, they assert that most research they survey fails to achieve real-time information queries, and lacks actuation elements.

\section{Real-time Requirements}\label{realtime}
As deployed IoT sensors become the standard for applications such as periodic monitoring of temperature and air quality in buildings, we propose that the next iteration of connected environments infrastructure focus around asynchronous real-time systems capable of collecting, analysing, predicting and adapting to change in a timely manner. There is a lack of attention paid in existing work to timeliness in such systems, perhaps due to the relatively limited scale of IoT deployments considered~\cite{dave2018framework,tang2019review}. We believe that the importance of event-driven real-time platform is thus insufficiently emphasised. We next clarify why this real-time component is essential, and how novel data flow and collection methods can be used to minimise response time.

\subsection{Key Concepts and Data Types}
A real-time focused system architecture is important because connected environments software need event-driven data to provide responsiveness without introducing arbitrary programmer-determined delays. Instead of being solely reliant on data querying using APIs, our proposed system architecture also enables instantaneous data flow, allowing for rapid event detection. By implementing this architecture we can overcome some of the hard limits related to IoT sensor deployment and BMS, such as the dependence on periodic data querying and static BIM systems.

In this section we outline key concepts related to an effective real-time data management platform for buildings. We distinguish different data types flowing through the platform, whilst simultaneously declaring that all data be treated as spatio-temporal data. We consider building data as spatial data, while all time-series sensor readings are temporal data, split into periodic and event-based data.

Many BIM systems assume spatial data is static by ignoring building metadata changes once construction is complete. We contend that this is false: buildings change over their life-cycle and so does their spatial data, either in the form of changes to floor plans or movement of building contents. A proper real-time platform integrating sensors, BMS and BIM must have the capacity to update spatial data properties as they change. This is particularly important in cases where building equipment often changes location, e.g.,~hospital beds, as coordinate data becomes (perhaps low-rate) real-time data subject to stream processing.

\subsection{Status Reading vs Events}
One of the fundamental challenges in real-time software platforms for sensor data is the lack of distinction between event data and periodic readings. This issue arises with the off-the-shelf sensor software which is often based on periodic status reporting rather than publish-subscribe and stream processing architectures.

While periodic readings are useful for sensor status reporting or sensing data that changes very gradually (e.g.,~air humidity), a platform solely based on periodic queries cannot be defined as real-time due to arbitrary blackout time-periods where no data is reported. For this reason, we believe a proper real-time architecture must focus on the concepts of \emph{timeliness} and \emph{events}, where messages are sent instantaneously after readings indicate a significant change in state.

\begin{figure}
  \centering
  \includegraphics[width=\linewidth]{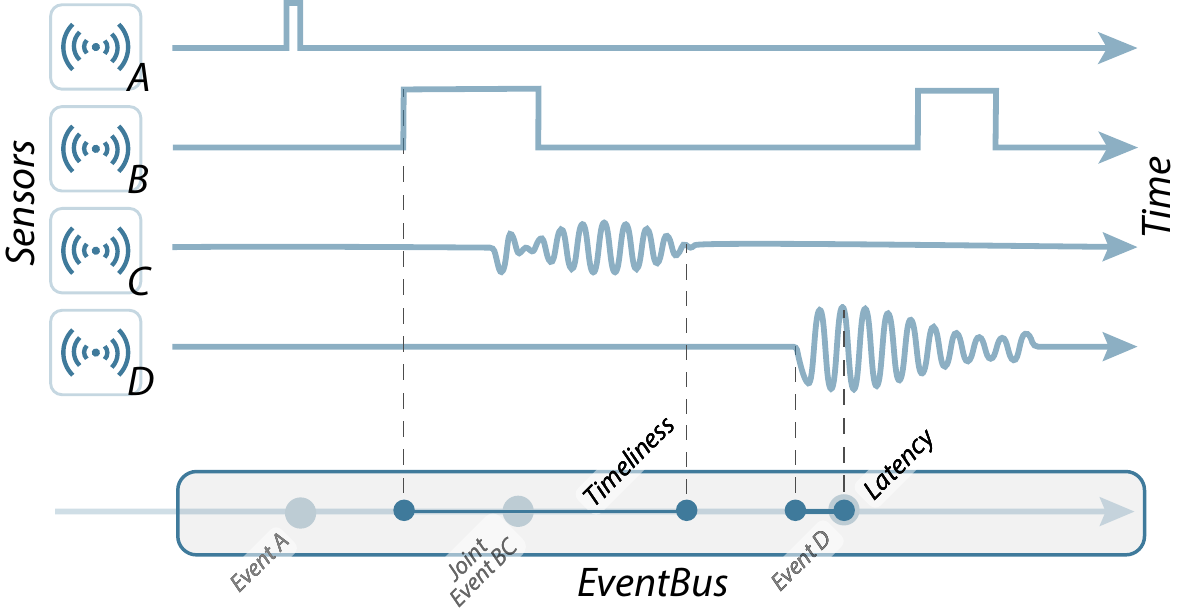}
  \caption{\label{timeliness_events} Concepts of Timeliness and Latency in events from spatio-temporal data. }
\end{figure}
\begin{figure}
  \centering
  \includegraphics[width=\linewidth]{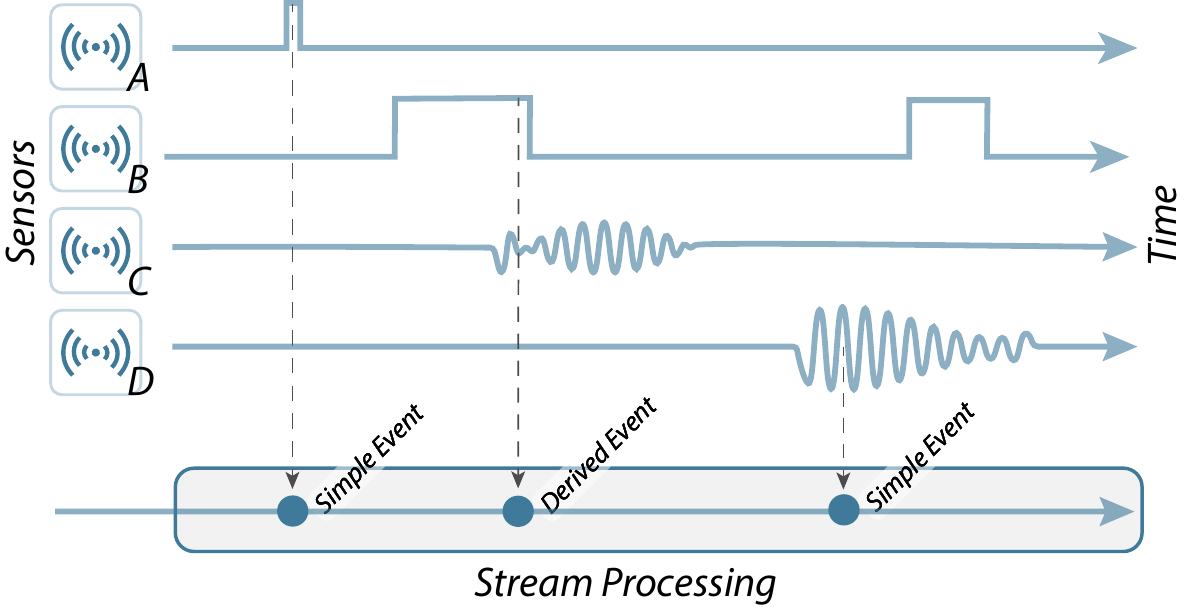}
  \caption{\label{simple_derived_events}Simple and Derived events from spatio-temporal data, illustrated by different types of sensor readings. }
\end{figure}

We define timeliness as a characteristic of an event, in which it is appropriate to act on changed readings, e.g. the duration from the beginning of the end of an event. In other words, it is a time-frame during which our event is at its most relevant. While for some events, e.g. interrupt-triggered sensor readings, latency and timeliness are almost identical time-wise, however, for more sophisticated events that follow a sequence of multiple sensor reading, minimising latency becomes difficult due to uncertainty.  As a result, we propose that every incoming event comes in with a probability value, determining the how likely the event-recognition is to be true.

Furthermore, the focus on events rather than raw data reporting allows for a flexible and easily scalable approach to data collection by having the ability to group raw sensor events into more informative, less frequent events. As the number of sensors increases, events can occur due to a single sensor being triggered, or as a result of a sequence of triggers passed down from multiple sensors. We distinguish two such categories of events: \emph{Simple} and \emph{Derived}.

\textbf{Event types}. Depicted in Figure~\ref{simple_derived_events}, we derive events based on their readings, e.g.,~when sensor readings reach a specific threshold or an interrupt occurs, for Sensors~A and~D. After a trigger occurs, the sensor sends an event to the real-time platform. We define such incoming data as \emph{Simple Events}. Sensors~B and~C are programmed to follow event detection based on a sequence of readings, e.g.,~sensor~B’s activation being followed by sensor~C's. In such cases, events are detected as a combination of sensor readings and we consider these \emph{Derived Events}.

Derived events are only possible when the timeliness of individual sensor readings overlap (Figure~\ref{timeliness_events}). Our platform permits us to look for such sequences of individual asynchronous events that result in more complex events being tracked, thus creating a richer building information environment.

Our intent is not to collect and solely rely on periodic readings and then send the data back to the platform, but rather to use stream processing to detect derived events as they happen. Real-time stream processing allows for data to be received and analysed instantly, therefore we are able not only to recognize events from a single sensor but also from a combination of multiple incoming data sources. This event-focused approach to real-time system architecture allows for non-blocking data processing with minimal additional latency and minimal network load.

\subsection{Programmatic Access to BMS}
In addition to event detection, another critical aspect of smart building infrastructure is the BMS-in-the-loop integration for building technical systems monitoring and actuation. While IoT sensor deployment is instrumental in smart buildings, programmatic access to BMS software for real-time usage suffers from the same lack of ontological standards that hinders deeper BIM-BMS integration.

The combination of a real-time platform that has access to both BMS and in-building sensors then allows for timely decision making by collecting, analysing, predicting and adapting to the sensed environment. As it stands, BMS software is limited to low granularity data (e.g.,~electricity consumption on a per-floor basis) and closed control loop mechanisms that would benefit from richer IoT data sources. While there are products on the market that claim to augment BMS capabilities by implementing MQTT drivers and deeper IoT integration~\cite{sip_platform}, the benefits of such platforms remain to be seen. In addition to well-known building automation benefits (e.g.,~decreased energy usage or HVAC optimisations), deeper BMS integration with the BIM software and IoT could be particularly useful in  \one~detecting accidents and emergencies,  \two~microgrid energy management,  \three~domain-specific applications.

\one~\emph{Accident and emergency management}. The use of a real time platform for accident and emergency response can form the basis for effective immediate risk management.
Broader IoT integration within BMS would play a critical role in recognising such anomalies in two ways. First, deployed sensors would be capable of detecting changes and anomalies in the behaviour of critical assets in real-time, avoiding further damage e.g., by monitoring pump vibrations and sensing any deviation from what is observed to be normal. Furthermore, the use of real time event recognition could be crucial in recognising gas or oil leaks in industrial settings where such accidents could be handled before becoming serious~\cite{muralidharaair}. Second, it would permit a general system-wide tracking across numerous subsystems deployed in buildings e.g.,~HVAC. A real-time BMS with access to highly localised sensor data would be especially useful in large scale infrastructure projects e.g., airports where numerous subsystems create a complex whole. Therefore, the resulting outcome becomes not only the provision of time-critical response systems but also a generally lowered complexity that makes large infrastructure projects easier to manage.

\two~\emph{Microgrid energy management}. Event-based occupancy tracking from a multitude of sensors deployed in the built environment allows for more efficient energy usage. Having the ability to combine sensor readings coming from environmental factors, like CO\textsubscript{2} concentration~\cite{franco2020measurement} and presence-detection indoors~\cite{zhang2019domain} presents numerous possibilities for heating optimisation and electricity consumption on a finer granularity a per room basis~\cite{perera2016estimation,zhang2019cooperative}.

\three~\emph{Domain-specific use cases}. Real time event detection can benefit built environments where building inhabitants are at an increased risk, e.g. in hospitals, for patient tracking and fall detection in wards~\cite{fan2017deep, gao2019real}. Similarly, in a smart home context, real time accident recognition could help the elderly population receive timely help if unable to live independently~\cite{tapia2004activity,alam2019besi}. Lastly, real time-based people density estimation to avoid overcrowding and violent behaviour detection in crowds can be useful for managing people flow in busy areas like stations or airports and stopping crowd crushes~\cite{niu2004human,hassner2012violent}.

\section{Design}
\label{design}
The core of our contribution is our real-time system, the Adaptive City Platform (ACP), as well as the accompanying example application showing our data pipeline in a real-world scenario. We connected our platform to sensors deployed in a non-BIM native building. The platform uses an ontology that we created to illustrate how real-time mechanisms allow for more versatility in BIM-IoT fusion.

\begin{figure}
  \centering
  \includegraphics[width=\linewidth]{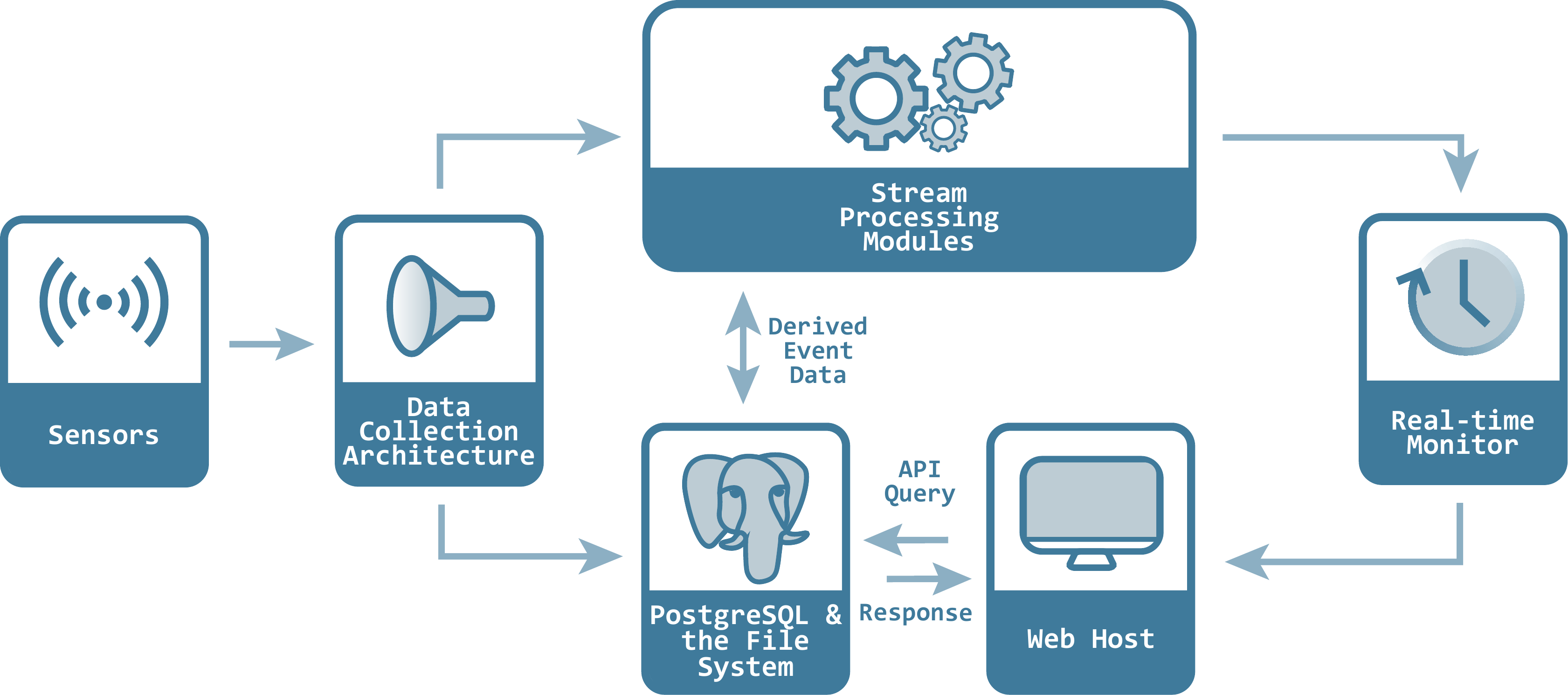}
  \caption{\label{sysarch}Adaptive City Platform system architecture.}
\end{figure}

Figure~\ref{sysarch} depicts the overall system. The ACP accepts real-time data flow from in-building sensors over the MQTT protocol. The back-end consists of a non-blocking real-time platform with a set of APIs accessing the file system and a PostgreSQL database.  Any incoming data is passed through our real-time platform where it is restructured, stored, and ultimately propagated to our front-end visualisation. The collected data is saved in a PostgreSQL database and the file system along with the BIM metadata.

We next describe our real-time platform architecture, APIs, and the front-end interface.

\subsection{Storage Structure}

In contrast to traditional XML based schemata used in BIM and other open data models like IndoorGML~\cite{teo2017extraction} or Brick~\cite{balaji2018brick}, the data inside the Adaptive City Platform is propagated using JSON objects. There were several reasons for this. First, it  enables the data to be easily readable  to humans and provides numerous benefits when working on both front-end and back-end languages~\cite{inbook}. Secondly, we envision the data flowing through the platform to utilise a Linked Data system, and we found JSON-LD to be the most compliant with our needs.

We annotate our data with auxiliary metadata properties, such as unique name identifiers, building descriptions and location data. This is done by appending metadata properties to the propagated data rather than replacing them, resulting in self-contained data structures that do not require additional API queries to be contextualised in use. For example, a sensor data object does not rely on querying the BIM to obtain information on the sensor's location, because that data is already embedded within the sensor metadata.

Our data structures can generally be separated into three distinct types: \one~building data, \two~sensor metadata and \three~sensor readings. This separation provides flexibility in data processing, as well as making it more usable for visualisation.Finally, sensor readings are stored in a data repository, where the data is sorted by date, as well as saved individually for each sensor and then sorted by date. While the data is duplicated this way, it does become useful when working with API queries that ask for data over a time period rather than from a specific sensor. Since we focused on a single building deployment, we found that saving our data in the file system worked sufficiently well, however, with a bigger deployment the use of a different data repository, e.g.  a SQL database, would be a viable option instead.

\subsection{Platform Modules}

The ACP real-time platform is based on the underlying architecture of the SmartCambridge framework~\cite{smartcambridge} which is implemented using Eclipse Vert.x~\cite{clement2015vert}, to allow for real-time event-based data processing. The platform consists of multiple Vert.x modules termed \emph{verticles} responsible for asynchronous, non-blocking data traversal. Figure~\ref{sysarch} illustrates these modules as Data Collection Architecture, Stream Processing, and Real-Time Monitor (RTMonitor) blocks.

\textbf{Data Collection Architecture.} Incoming sensor data is captured by our Data Collection verticles that listen for incoming messages over MQTT. After a message is received, the modules timestamp binary sensor data, archive it in the file system and further propagate sensor readings down the platform as JSON files. Using the SmartCambridge framework allowed us to reach an average latency of 160ms from sensors sending the data to its use in the sample application.

\textbf{Stream Processing Modules.} Stream Processing is comprised of modules that further parse the received JSON files. These modules decode and write the data to the file system as well as inserting relevant metadata in the database and passing data to the RTMonitor for low-latency data visualisation. Stream Processing modules are versatile because they are able to sort the incoming data and separate it into different directories per event basis for future API use.

\textbf{RTMonitor.} The RTMonitor is a module allowing client web pages to issue subscriptions to the Stream Processing verticles and receive updates via websockets on the specified URI the moment sensor readings are passed down the pipeline. The RTMonitor then manages these subscriptions using a token-based system, where users can issue subscriptions by specifying the sensor data they would like to receive.


However, while the RTMonitor is an integral part of the real-time platform, we also use regular API queries to fetch historical data and BIM-related metadata from our database.

\subsection{APIs}
We created four APIs to fetch the sensor metadata, sensor readings, BIM metadata and rendered SVG BIM data. These API endpoints are used in our example application to provide the scaffolding that permits the non-blocking real-time data flow straight to the front-end visualisation.

\begin{table}
  \centering
  \begin{tabular}{l|l}
    \textit{/bim} & returns building metadata\\  \hline
    \textit{/space} & returns rendered SVG objects based on the BIM data \\ \hline
    \textit{/sensors} & returns sensor metadata\\ \hline
    \textit{/readings} & returns sensor readings \\ \hline
  \end{tabular}
  \caption{\label{api_desc}API description.}
\end{table}

\begin{figure}
  \centering
  \includegraphics[width=\linewidth]{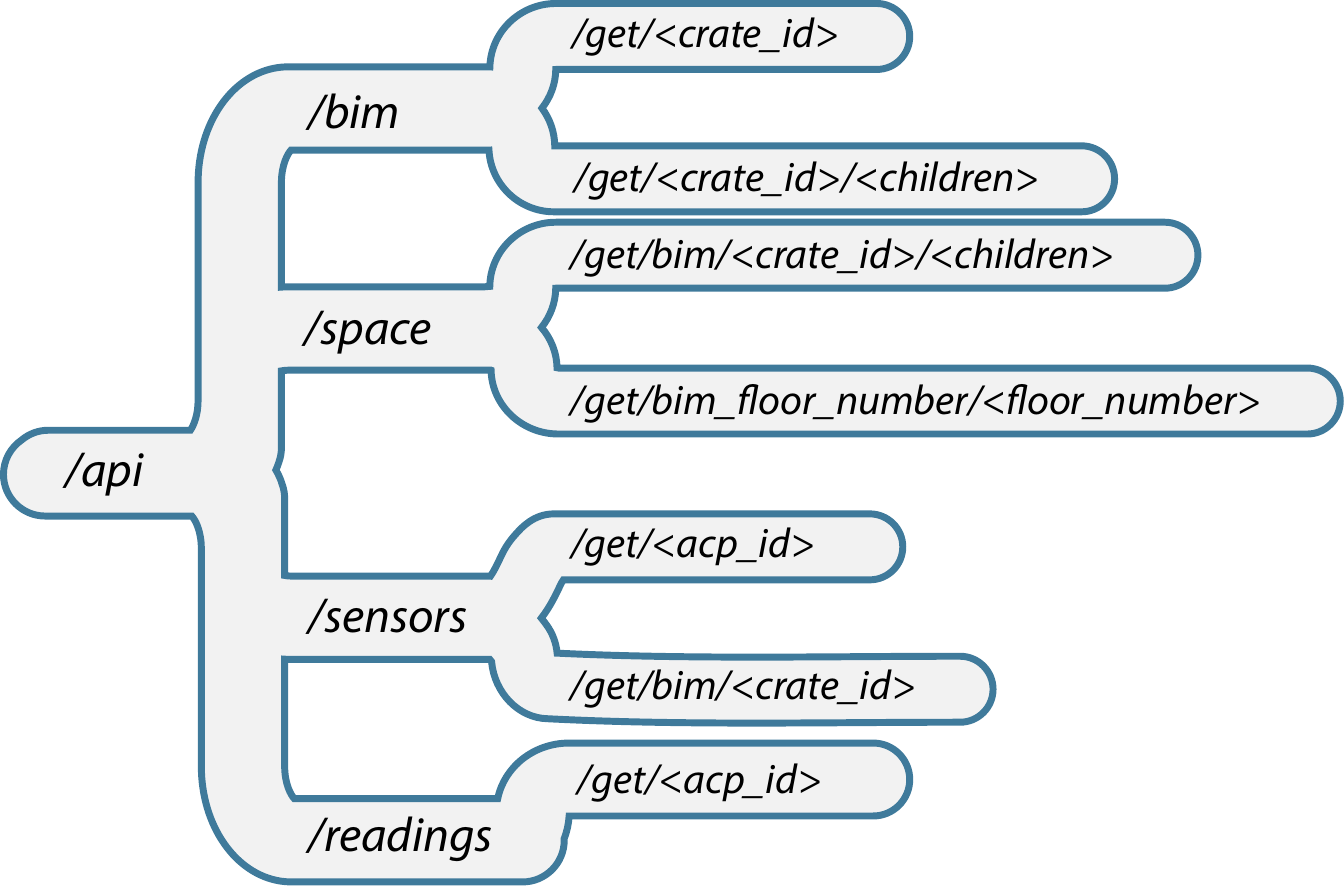}
  \caption{\label{api_struct}API structure.}
\end{figure}

The front-end is maintained using the API endpoints that interface with the PostgreSQL database and the file system to query the BIM and sensor data. Three of the four API endpoints return JSON-formatted historical data that is easily manipulated by JavaScript or Python. The \textit{/space} API endpoint is unique in that it returns nested XML SVG objects that are rendered on screen for the building visualisation. The diagram shown in Figure~\ref{api_struct} and the Table~\ref{api_desc} describes the API in more detail.

\subsection{Building Reference Data Platform}
\label{building_data_structs}

In order to evaluate the Adaptive City Platform, we deployed and tested IoT sensors in a non-BIM native building, that did not have any Revit files, other than 2D floor plans.
Initially we considered adopting a BRICK ontology for the building, however, after testing it we decided to craft a JSON-LD compliant proprietary ontology with the ability to be converted to BRICK. We did this for the following reasons.

First, BRICK’s focus is to capture information on building operations and improve application portability -- not to create another building design model~\cite{balaji2018brick}, like the IFC standard. This was problematic as our system was deployed in a non-BIM native building that lacked any IFC files for contextual visualisation. Thus, the ability to embed precise location coordinates and boundary data, in addition to relational information, was important for us. While achieving this using the BRICK ontology was technically feasible by saving the additional properties as literals, we found it to be impractical. The impracticality was caused by our metadata properties existing as dictionaries, and hence saving them as strings would \textit{(a)} reduce human readability and \textit{(b)} slower data retrieval time, as string literals would have to be converted back to objects.

Second, we found that including additional metadata properties like timestamps for recognising patterns were key. For our application area, spatio-temporal data is highly relevant, but BRICK mostly ignores such information by disregarding time, making it unsuitable for true real time system applications.

Third, we found that our existing JSON (and further JSON-LD) data structures, whilst providing more flexibility than RDFa used by BRICK, could be easily converted to BRICK ontology, providing the possibility of portability when needed.
Therefore, we crafted a sample ontology to reference our building and sensor metadata in the platform.

For our building reference data we used a universal naming convention for buildings, floors and rooms called \textit{crates}. A \textit{crate} is an object (however, usually a building or part of a building) with a defined enclosed boundary that denotes its perimeter. \textit{Crates} can also be nested by defining the outer \textit{crate} as a \textit{parent crate}. For example, we can define a \textit{crate} that denotes a building, and within that building we have other \textit{crates} that are its children -- e.g.,~floors, that in return also have their own children -- e.g.,~\textit{room crates}. This parent-child relation in our building data allows for hierarchical data management that is useful for efficient API queries.

\begin{figure}
  \centering
  \includegraphics[width=.8\linewidth]{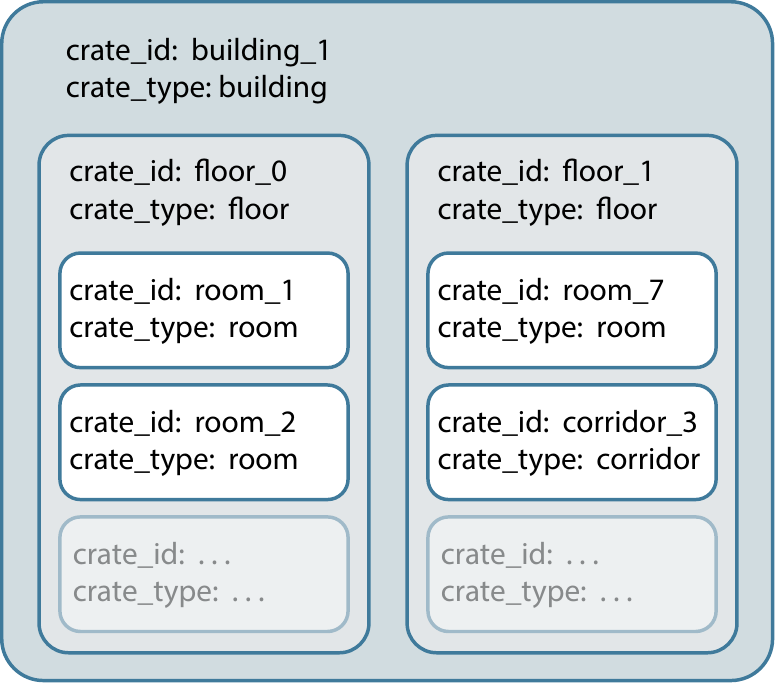}
  \caption{\label{crate_struct}\textit{Crate} structure illustrating the parent-child relation and object types.}
\end{figure}

We use the nested tree-like structure in our BIM database to retrieve data associated with our building. For this reason, the API call \textit{/bim/get/<crate\_id>/<children>} has the optional \textit{<children>} property that allows us to specify if any children should be included for that particular crate.

Additionally, every \textit{crate} has a \textit{ crate\_type} property to select specific object types, e.g.,~\textit{floor}, \textit{building}, etc., as shown in Figure~\ref{crate_struct}. Further metadata associated with \textit{crates} includes location, boundary and crate type, as well as a UNIX timestamp that indicates the date any information had been updated. Overall, using this metadata structure permits us to index our objects in the PostgreSQL database using any of the properties defined below to allow for fast and flexible metadata retrieval and data management. A complete example of the BIM metadata is given in Table~\ref{BIM_metadata}, and a sample response to the BIM API query is shown in Listing~\ref{crate-json-example}.

\begin{table*}
  \begin{tabular}{l|l|l|l|l}
    \textit{crate\_id} & \textit{parent\_crate\_id} & \textit{crate\_type} & \textit{location} & \textit{boundary} \\ \hline
    WGB &
          - &
              “building” &
                           \begin{tabular}[c]{@{}l@{}}\{   "system":"GPS", \\    "acp\_lat": -27.116667, \\    "acp\_lng":-109.366667, \\    "acp\_alt":0.0 \}\end{tabular} &
                                                                                                                                                                              \begin{tabular}[c]{@{}l@{}}\{\\   "system":"WGB",\\   "boundary":{[} {[}0,0{]},{[}..,...{]},{[}..,..{]},{[}245,56{]} {]}\\ \}\end{tabular} \\ \hline
    GF &
         WGB &
               “floor” &
                         \begin{tabular}[c]{@{}l@{}}\{   "system":"WGB", \\    "x":36.5, \\    "y":39, \\    "f":0, \\    "zf":0 \}\end{tabular} &
                                                                                                                                                   \begin{tabular}[c]{@{}l@{}}\{\\   "system":"WGB",\\   "boundary":{[} {[}0,0{]},{[}..,..{]},{[}..,..{]},{[}245,56{]} {]}\\ \}\end{tabular} \\ \hline
    FE11 &
           FF &
                “room” &
                         \begin{tabular}[c]{@{}l@{}}\{   "system":"WGB", \\    "x":22.06, \\    "y":34.67, \\    "f":1, \\    "zf":0 \}\end{tabular} &
                                                                                                                                                       \begin{tabular}[c]{@{}l@{}}\{\\   "system":"WGB",\\   "boundary": {[} {[}0,0{]},{[}0,78{]},{[}73,78{]},{[}73,0{]} {]}\\ \}\end{tabular} \\ \hline
  \end{tabular}
  \caption{\label{BIM_metadata}BIM metadata example, extracted from our database, showing how data can be indexed based on range of properties.}
\end{table*}

\begin{listing}
  \small\begin{minted}[frame=none,
    framesep=1mm,
    linenos=false,
    xleftmargin=9pt,
    tabsize=4]{js}
    {
      "crate_id": "FE11",
      "crate_type": "building",
      "acp_ts": "1589469825.165538",
      "long-name": "Computer Science Department",
      "description": "Crate Description",
      "acp_boundary": "[
      [0,0]  ,[0,78],
      [73,78],[73,0]
      ]",
      "parent_crate_id": "West Campus",
      "acp_location":{
        "f": 1,
        "x": 22.06,
        "y": 34.67,
        "z": 0,
        "system": "WGB"
      }
    }

  \end{minted}
  \caption{\label{crate-json-example}Crate object metadata, received after querying the API endpoint \textit{/bim/get/FE11}.}
\end{listing}

\subsection{Sensor Reference Data Platform}
\label{sensor_data_structs}

\begin{table}
  \centering
  \begin{tabular}{p{2cm}|p{5.6cm}}
    \hline
    \textit{key}               & \textit{definition}                                                                                      \\
    \hline
    acp\_id           & sensor identifier, globally unique e.g.,~elsys-eye-049876.                                       \\
    \hline
    acp\_ts           & epoch timestamp most relevant to data reading or event, e.g.,~"1586461606.465372".               \\
    \hline
    acp\_type         & sensor type, determines data format e.g.,~elsys-eye.                                             \\
    \hline
    acp\_event\_value & qualifier or data reading for event, e.g.,~open.                                                 \\
    \hline
    acp\_event        & event type, for time-stamped events, e.g.,~openclose.                                            \\
    \hline
    acp\_location     & location using a custom coordinate system e.g.,~\{ "system": "WGB", "x": 12, "y": 45, "f": 1 \}.  \\
    \hline
    acp\_confidence   & a value 0-1 indicating the reliability of the sensor reading.                                   \\
    \hline
  \end{tabular}
  \caption{\label{sensor_props}Sensor metadata properties.}
\end{table}

\begin{listing}
  \small\begin{minted}[frame=none,
    framesep=1mm,
    linenos=false,
    xleftmargin=9pt,
    tabsize=4]{js}

    {
      "acp_id": "elsys-co2-041ba9",
      "acp_ts": "1589469979.861816",
      "type": "co2",
      "owner": "ijl20",
      "source": "mqtt_ttn",
      "features": "co2, humidity,
      light, motion,
      temperature, vdd",
      "acp_location": {
        "system":"GPS",
        "acp_alt": 10,
        "acp_lat": -27.116667,
        "acp_lng": -109.366667,
        "parent_crate_id": "FE11"
      }
    }
  \end{minted}
  \caption{\label{sensor-json-example}Sensor object metadata, received after querying the API endpoint \textit{/sensors/get/elsys-co2-041ba9}.}
\end{listing}

Similarly to \textit{crate\_id}, sensors also have unique identifiers, the \textit{acp\_id} property, defined using the manufacturer’s name and a six character-long unique identifier. Since different types of sensors often have varying attributes, our sensor metadata table is defined by two columns containing: \one~the \textit{acp\_id} identifiers and \two~another containing the other auxiliary properties. Such auxiliary metadata includes information on sensor properties like the location and type. Complete metadata description can be found in Table~\ref{sensor_props}, as well as a queried sample sensor metadata object from \textit{/sensors/get/<acp\_id>} in Listing~\ref{sensor-json-example}.

Finally, we expanded the sensor API  to be capable of retrieving sensor metadata by their physical deployment location, specifying the parent crate in the API call \textit{/sensors/bim/get/<crate\_id>}.

\subsection{Historical Sensor Data Platform}

We attach UNIX timestamps to all metadata and incoming sensor readings as soon as they enter the ACP. Depending on the type of data received, JSON packets further propagate to the PostgreSQL database or the file system, where they become historical data. Historical sensor readings can be fetched by querying the API with a sensor’s \textit{acp\_id} property. The API call then returns the following JSON file from the most recent entry in the database, as shown in Listing~\ref{reading-json-example}.

\begin{listing}
  \small\begin{minted}[frame=none,
    framesep=1mm,
    linenos=false,
    xleftmargin=9pt,
    tabsize=4]{js}

    {
      "acp_id": "elsys-co2-041ba9",
      "acp_ts": "1589469979.861816",
      "features": {
        "co2": 415,
        "device": "elsys_co2",
        "humidity": 36,
        "light": 0,
        "motion": 2,
        "temperature": 15.3,
        "vdd": 3659
      }
    }

  \end{minted}
  \caption{\label{reading-json-example}Sensor reading data received after querying the API endpoint \emph{/readings/get/elsys-co2-041ba9}.}
\end{listing}

\subsection{Spatial Coordinates Data Platform}

In both building and sensor metadata we refer to our spatial coordinate system as \textit{acp\_location}. As the ACP needs to provide data for both relative (in-building) and global (WGS84) positioning references for sensors and \textit{crates} alike, we introduce three parallel location reference systems.

\textbf{Global}.  The definitive common reference system constituting of latitude, longitude and altitude. The global system is necessary for outdoor sensors as latitude and longitude coordinate system is used while interacting with the \textit{site template} view in the sample application (Figure~\ref{all_bim_templates}A). We define this in our BIM model by setting the \textit{acp\_location} parameter as \textit{GPS}, shown in Table~\ref{BIM_metadata}.

\textbf{In-building coordinates}. We use a spatial coordinate system unique to each building, typically when interacting with in-building floor plan or 3D views of sensors or data. Sensors that transmit their position in the building (particularly relevant for  sensors that move around) may use this system in their sensor data. We set in-building coordinates by specifying the \textit{acp\_location}’s  parameter as the  building’s name, as illustrated in Table~\ref{BIM_metadata} and Listing~\ref{crate-json-example}.

\textbf{Building object hierarchy}. Hierarchies are often used in BIM software, and in the ACP the hierarchy is defined with the \textit{parent\_crate} parameter. It is reasonable for a sensor (or other monitored device) to be recorded as being in location based on a \textit{crate} i.e.~a room/office, which relates to the BIM data structured as \textit{site→building→floor→room→window}, etc. This hierarchy is often natively used when collating or browsing in-building information, such as electricity consumption in specific rooms or floors.

By annotating our metadata with a combination of these three location systems, we achieve high flexibility to manage and visualise the BIM and sensor data in multiple ways. This is exemplified in the following section where we showcase how the real-time platform can be effectively used in a real-world scenario.

\subsection{Visualisation}

\begin{figure*}
  \centerline{\includegraphics[width=\linewidth]{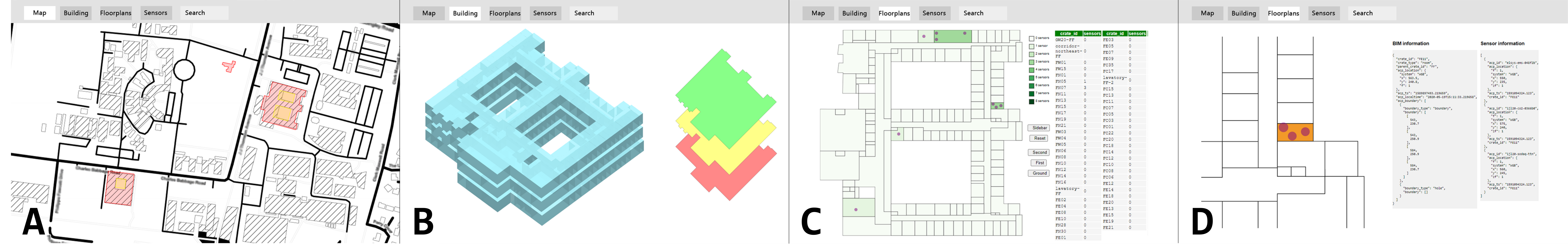}}
  \caption{\label{all_bim_templates} The 4 templates used to show BIM data on different scales: (A)~Site, (B)~Building, (C)~Floor, (D)~Floorspace.}
\end{figure*}

To best illustrate how the ACP can perform in the wild, we conceived a data visualisation application to monitor and visualise the BIM-IoT fusion data. The visualisation acts as an example of how we can develop software based around our platform and APIs to facilitate the BMS and IoT integration.

\begin{figure}
  \centering
  \includegraphics[width=\linewidth]{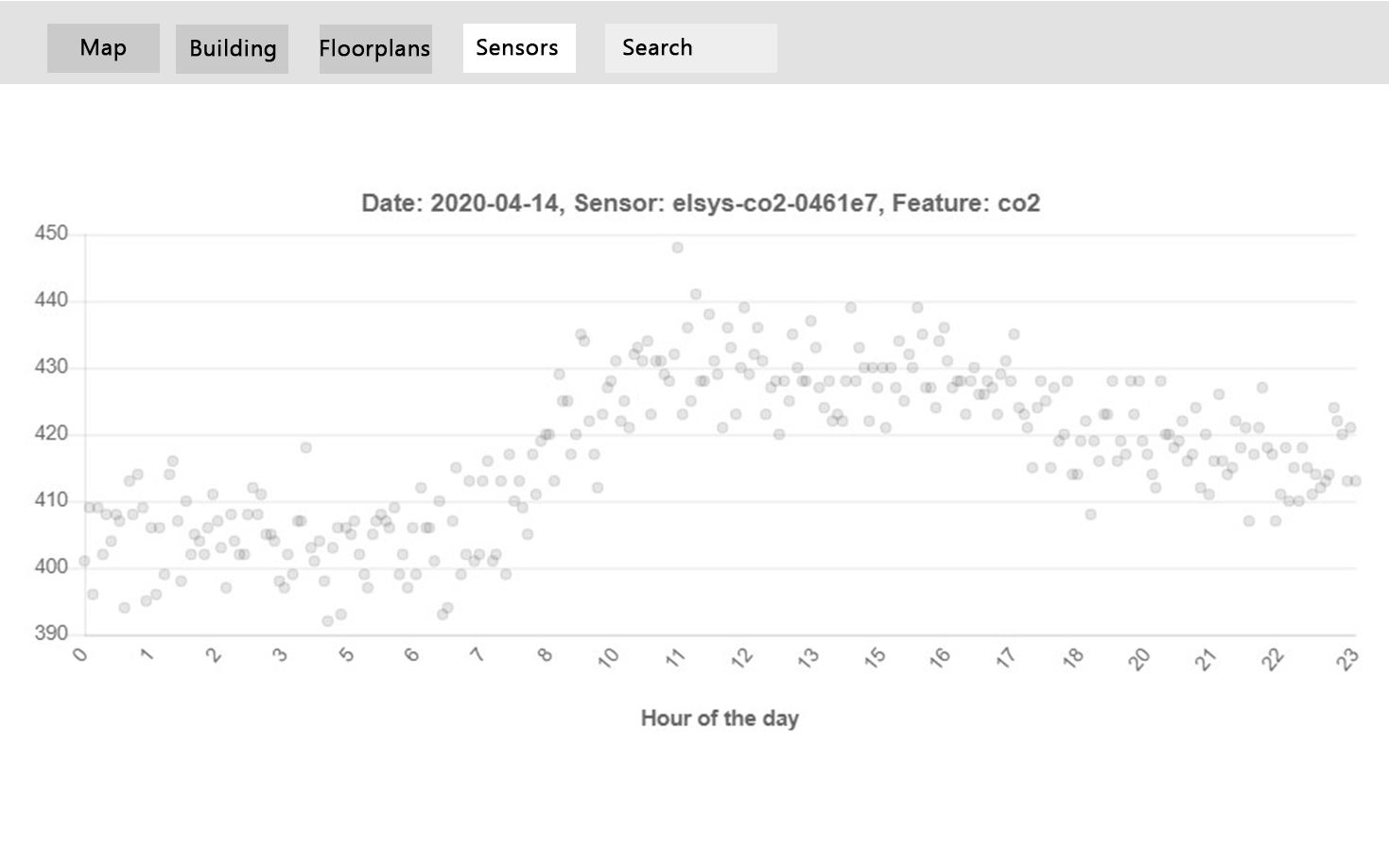}
  \caption{\label{sensor_template}A template used to visualise a day's worth of historic sensor data for one of the deployed CO\textsubscript{2} sensors.}
\end{figure}

The front-end consists of five templates showing the BIM-IoT fusion at different levels. We use  multiple templates for our data visualisation, allowing us to select the granularity at which the information is displayed. The templates are loaded in a hierarchical order, where the first template (Figure~\ref{all_bim_templates}A) represents the site view, while the last one (Figure~\ref{all_bim_templates}D) shows the \textit{crate-level} view, followed by a sensor readings template (Figure~\ref{sensor_template}) with time-series plots.

\subsubsection{Site Template}
The site-level visualisation (Figure~\ref{all_bim_templates}A) displays buildings and sensors in aggregated groups on the campus. The map is rendered using Leaflet and OpenStreetMap~\cite{OpenStreetMap}, an open-source JavaScript library for interactive maps.

We query our database to return the boundary data for the buildings we would like to render on the screen. After receiving the boundary coordinates in latitude and longitude we then render polygons over the original building locations on the map. Finally, users are able to navigate to the other templates by clicking on individual buildings.

\subsubsection{Building Template}
In order to render the building-level template we query the API to return individual floor and room boundaries based on the information we have in the BIM database. After fetching the boundary data, we then render the building in 3D using the Three.js library~\cite{noauthororeditor2015threejs}.

The building-level template (Figure~\ref{all_bim_templates}B) allows for a general building view to inspect individual crates, or proceed to any of the floor plan templates by clicking on the floor icon.

\subsubsection{Floor Plan Template}
The floor-level template (Figure~\ref{all_bim_templates}C) has several important features. Primarily, it has the ability to display heat-maps of sensor readings, as well as the most recent sensor data. Users can inspect sensor readings by hovering over the sensor icon on the floor plan. Upon doing so, an API call is executed fetching the most recent data for the particular sensor in question.

We query the \textit{/space/get\_bim\_floor\_number/<floor>} API endpoint to acquire all crates that are on the queried floor. The API returns an SVG object along with the metadata necessary to render the floors and individual rooms. We then use D3.js~\cite{d3js} to render the SVG on screen and make it interactive.

With this approach we eliminate the need to possess static SVG files in the file system, and instead rely on our predefined BIM data structures that can be easily updated and generated into SVG files on demand.

\begin{listing}
  \small\begin{minted}[frame=none,
    framesep=1mm,
    linenos=false,
    xleftmargin=9pt,
    tabsize=4]{xml}

    <g>
      <polygon
        id='FE11'
        data-crate_type='room'
        data-parent_crate='FF'
        data-floor_number='1'
        points='362.8,0 362.8,40.3 482.1,40.3 482.1,0'>
        <title>
          FE11
        </title>
      </polygon>
    </g>

  \end{minted}
  \caption{\label{SVG-example}A generated \textit{crate} SVG sample on the floor-level visualisation. We embed additional metadata to SVG objects using the HTML \textit{data-*} attribute.}
\end{listing}

Users proceed to enter the floorspace template by clicking on a crate on the floor-level template.

\subsubsection{Floorspace Template}
This final BIM-generated template (Figure~\ref{all_bim_templates}D) is used to display the finest granularity spatial data, including the \textit{crate}’s BIM description and metadata for all the sensors deployed in that crate. The selected \textit{crate} of type \textit{room} is zoomed in on, allowing to more accurately display the sensor location.

\subsubsection{Sensor Template}
As users travel down the spatial hierarchy of the web application, the data becomes increasingly more granular, allowing detailed time-series data analysis. The sensor-level template is accessed by clicking on a sensor on either the floor plan or floor space views. The user is then redirected to the sensor template where detailed spatio-temporal time-series data is shown, as illustrated in Figure~\ref{sensor_template}. This template allows to visualise the data provided by every in-building sensor. Users are able to select the timeframe they wish to inspect. The visualisation changes in real-time as new data readings propagate through the platform.

\section{Discussion}
\label{discussion}

During the development of the ACP and the accompanying visualisations, we identified four major themes associated with BIM-IoT fusion: \one~scalability, \two~sensor-derived constraints, \three~BMS actuation and \four~privacy.

\subsection{Scalability}
While the ACP does provide real-time sensor information, it remains a single building deployment with a limited scope. We anticipate using our modular architecture to add additional buildings simply by appending the BIM files to the current PostgreSQL database.

One challenge in doing so is to acquire the BIM data in the first place. In our example we used a non BIM-native building and as a result had no access to 3D building models and other key BIM data. Instead we parsed the SVG floor plans for our building to extract the boundary data for individual crates. While such CAD file parsing could also be adapted to other 2D-friendly formats such as DXF or DWG, the differences between the way building data is stored remains an issue. As use of BIM in the design phase of buildings increases, a potential solution would be to write a Revit plugin that would automatically generate and export the requested data in a suitable ontological standard~\cite{rasmussen2018madsholten}, or utilise IFC-BRICK converter~\cite{balaji2018brick}.

In terms of the back-end system performance, we have no reason to suspect that any further scalability efforts would affect the real time platform itself. Since our real-time architecture is based on the Vert.x system used in the SmartCambridge project~\cite{smartcambridge} that had $\sim$1000 sensors sending data every 20 seconds, we have encountered no slowdown with asynchronous modules processing data with low-latency.

\subsection{Sensor-produced constraints}
Event-driven data is crucial for responsiveness but the majority of  off-the-shelf sensors, with the exception of interrupt-based devices, are only capable of sending periodic readings, e.g.,~CO\textsubscript{2} every 5 minutes. Furthermore, such sensors are rarely reconfigurable to allow for dynamic data transmission based on user-defined threshold values,  making it impossible to  implement timely event-driven behaviour to track sensor readings in real-time. Finally, many sensors lack control over where data is delivered -- while we chose to rewrite factory sensor firmware in some cases, the complexity and the required man-hour efforts may not necessarily be worth the desired outcome.

Overall, sensor choice and their detailed capabilities plays a key role in determining the timeliness factor of any real-time platform, therefore considerable attention should be paid to what sensors are integrated within the IoT ecosystem.

\subsection{BMS Implementation and Actuation}
As research progresses towards the adaption of a universal ontological standard to incorporate BIM and BMS with IoT technologies, machine learning-based approaches to building automation and actuation will become more important. According to Tang et al, many current BIM-IoT fusion attempts fail to achieve the enclosed loop system capability by not enacting on actuation and only being capable of sensing~\cite{tang2019review}.

\subsubsection{BMS in-the-loop}
BMS implementation is often a non-trivial task due to siloed software systems, as well as some buildings not having actuation capabilities implemented and accessible by their BMS~\cite{becerik2012application}. While beyond the scope of this paper, such attempts to normalise BMS data are made through the creation of building ontologies, providing unified metadata schemas.

However, a substantial number buildings have neither BMSs nor BIM-enabled facilities management systems, and therefore lack basic programmatic access. While deploying IoT sensor would make such buildings ‘smarter’, the lack of actuation equipment means that we would not be able to reach the full potential of modern, BIM-native buildings.

While implementing actuation components in such buildings would likely be the next step after the deployment of sensors and machine learning tools, it would also mean a substantial financial commitment to AEC stakeholders. Nevertheless, it has been reported that building automation does indicate a substantial energy savings~\cite{zucker2012building,cascone2017ethical}, thus potentially making expenses cost-effective, even ignoring the benefits that such systems could bring in risk/disaster management and anomaly detection.

\subsubsection{Data Analysis, Prediction and Actuation}
Even though HVAC systems in buildings have been around for a considerable time and have shown positive results in increasing energy efficiency~\cite{zucker2012building}, the mass deployment of IoT sensors in the built environment will provide much more granular data feeds. This increase in the amount and detail of data is what drives the need for fast and efficient real-time platforms like the ACP, as they permit the analysis and decision making to be done on the spot, once the right machine learning tools are implemented in the data pipeline.

The ongoing COVID-19 pandemic prevented us progressing our deployment to the point where we had dense enough deployment of sensors and rich enough data from those sensors, building automation and data analytics remains a topic of great importance for future research.

\subsection{Privacy}
Privacy in IoT-backed smart buildings has grown in importance over the last decade~\cite{cascone2017ethical,agarwal2016toward,hui2017major}, but more work is needed as privacy issues remain a minefield and may hinder further technology adoption~\cite{hernandez2015safir}. It is important to note that while sensing in the presence of people in itself causes privacy concerns, real-time tracking further amplifies these concerns, as data integrity may not always be guaranteed~\cite{tang2019review}. Even though none of the sensors are used to track individuals directly, latent information may still cause ethical issues, as it may be possible to, e.g.,~infer information about office workers by analysing CO\textsubscript{2} data and electricity consumption~\cite{atlam2020iot}.

The primary concern about deploying sensors in the built environment has to do with the ownership of and the access to the collected data. Sensors are relatively easily locatable objects in the workplace and a possible source of anxiety due to the collection of data. Currently there are few IoT deployment guidelines covering office spaces and data ownership, other than the GDPR, focused on data collection and storage~\cite{voigt2017eu}. We envision a privacy-oriented system provided by a capability-based platform where people are given access to data based on a combination of factors such as their day-to-day proximity to sensors (you should have means to examine data collected about your behaviour), and role in the company (a building manager may need to see aggregate data for the whole building).

For example, people with sensors deployed in their offices would have the access and control of the high granularity (or frequency) time series data that they produce. However, such access would be limited to the rest of the building, providing a decreasing level of granularity (or frequency) across time and space domains. An example of this in practice could be the building facility manager having the access to entire building data on a weekly scale without specific information on individual offices.

In addition to personal privacy in the workplace, in a case study presented by Cascone et al, security of data was the second most common concern after privacy~\cite{cascone2017ethical}. The deployment of similar web applications to ours comes with a risk of data being leaked to third parties. Therefore, considerable care should be taken to secure both the BIM data, and the sensor data so that it would only be accessible to authorised personnel. The same capability-based platform could be employed in cases where third parties are given access to some of the low frequency, low resolution spatial data.

\section{Conclusion}
\label{concls}

The lack of standardisation in BIM-IoT fusion means that the AEC industry is unprepared for mass IoT sensor deployment in the built environment. Our contribution to research on BIM-IoT fusion has been to describe a high-level real-time software architecture for processing data, and to show how it can be used to visualise in-building sensor data with useful features for facility managers and building inhabitants alike.

As it stands, our real-time Adaptive City Platform is currently unique, playing a key role in the way the IoT sensors are integrated with BIM. By using the concept of events in capturing time-sensitive data in the ACP we were able to achieve low-latency stream processing capable of instantaneous updates to our in-house visualisation.

We have further shown how hierarchical data visualisation can be used to display the collected data at different levels of granularity. Furthermore, we have described how metadata is described and our data is stored, both buildings (\emph{crates}) and sensors.

While we have demonstrated that the ACP works, there are still substantial challenges in how to tackle mass sensor deployment and further data analytics. As workplaces and buildings eventually reopen as the current phase of the COVID-19 pandemic subsides, we anticipate being able to increase the density and spread of our deployment, enabling us to collect and process more and richer data to explore how the ACP can support prediction and actuation.


{
  \bibliographystyle{ACM-Reference-Format}
  \bibliography{buildsys}


\begin{thebibliography}{53}


\ifx \showCODEN    \undefined \def \showCODEN     #1{\unskip}     \fi
\ifx \showDOI      \undefined \def \showDOI       #1{#1}\fi
\ifx \showISBNx    \undefined \def \showISBNx     #1{\unskip}     \fi
\ifx \showISBNxiii \undefined \def \showISBNxiii  #1{\unskip}     \fi
\ifx \showISSN     \undefined \def \showISSN      #1{\unskip}     \fi
\ifx \showLCCN     \undefined \def \showLCCN      #1{\unskip}     \fi
\ifx \shownote     \undefined \def \shownote      #1{#1}          \fi
\ifx \showarticletitle \undefined \def \showarticletitle #1{#1}   \fi
\ifx \showURL      \undefined \def \showURL       {\relax}        \fi
\providecommand\bibfield[2]{#2}
\providecommand\bibinfo[2]{#2}
\providecommand\natexlab[1]{#1}
\providecommand\showeprint[2][]{arXiv:#2}

\bibitem[\protect\citeauthoryear{??}{mem}{2016}]%
        {memoori2018}
 \bibinfo{year}{2016}\natexlab{}.
\newblock \bibinfo{title}{The Internet of Things in Smart Commercial Buildings
  2016 to 2021}.
\newblock
\newblock
\urldef\tempurl%
\url{http://memoori.com/portfolio/internet-things-smart-commercial-buildings-2016-2021/}
\showURL{%
\tempurl}


\bibitem[\protect\citeauthoryear{Agarwal and Dey}{Agarwal and Dey}{2016}]%
        {agarwal2016toward}
\bibfield{author}{\bibinfo{person}{Yuvraj Agarwal} {and}
  \bibinfo{person}{Anind~K Dey}.} \bibinfo{year}{2016}\natexlab{}.
\newblock \showarticletitle{Toward Building a Safe, Secure, and Easy-to-Use
  Internet of Things Infrastructure.}
\newblock \bibinfo{journal}{\emph{IEEE Computer}} \bibinfo{volume}{49},
  \bibinfo{number}{4} (\bibinfo{year}{2016}), \bibinfo{pages}{88--91}.
\newblock


\bibitem[\protect\citeauthoryear{Alam, Homdee, Wolfe, Hayes, and Lach}{Alam
  et~al\mbox{.}}{2019}]%
        {alam2019besi}
\bibfield{author}{\bibinfo{person}{Ridwan Alam}, \bibinfo{person}{Nutta
  Homdee}, \bibinfo{person}{Sean Wolfe}, \bibinfo{person}{James Hayes}, {and}
  \bibinfo{person}{John Lach}.} \bibinfo{year}{2019}\natexlab{}.
\newblock \showarticletitle{Besi: behavior learning and tracking with wearable
  and in-home sensors-a dementia case-study}. In
  \bibinfo{booktitle}{\emph{Proceedings of the International Conference on
  Internet of Things Design and Implementation}}. \bibinfo{pages}{281--282}.
\newblock


\bibitem[\protect\citeauthoryear{Antonopoulou and Bryan}{Antonopoulou and
  Bryan}{2017}]%
        {antonopoulou2017bim}
\bibfield{author}{\bibinfo{person}{Sofia Antonopoulou} {and}
  \bibinfo{person}{Paul Bryan}.} \bibinfo{year}{2017}\natexlab{}.
\newblock \bibinfo{booktitle}{\emph{BIM for heritage: developing a historic
  building information model}}.
\newblock \bibinfo{publisher}{Historic England}.
\newblock


\bibitem[\protect\citeauthoryear{Atlam and Wills}{Atlam and Wills}{2020}]%
        {atlam2020iot}
\bibfield{author}{\bibinfo{person}{Hany~F Atlam} {and} \bibinfo{person}{Gary~B
  Wills}.} \bibinfo{year}{2020}\natexlab{}.
\newblock \showarticletitle{IoT security, privacy, safety and ethics}.
\newblock In \bibinfo{booktitle}{\emph{Digital Twin Technologies and Smart
  Cities}}. \bibinfo{publisher}{Springer}, \bibinfo{pages}{123--149}.
\newblock


\bibitem[\protect\citeauthoryear{Azhar}{Azhar}{2011}]%
        {azhar2011building}
\bibfield{author}{\bibinfo{person}{Salman Azhar}.}
  \bibinfo{year}{2011}\natexlab{}.
\newblock \showarticletitle{Building information modeling (BIM): Trends,
  benefits, risks, and challenges for the AEC industry}.
\newblock \bibinfo{journal}{\emph{Leadership and management in engineering}}
  \bibinfo{volume}{11}, \bibinfo{number}{3} (\bibinfo{year}{2011}),
  \bibinfo{pages}{241--252}.
\newblock


\bibitem[\protect\citeauthoryear{Azhar, Hein, and Sketo}{Azhar
  et~al\mbox{.}}{2008}]%
        {azhar2008building}
\bibfield{author}{\bibinfo{person}{Salman Azhar}, \bibinfo{person}{Michael
  Hein}, {and} \bibinfo{person}{Blake Sketo}.} \bibinfo{year}{2008}\natexlab{}.
\newblock \showarticletitle{Building information modeling (BIM): benefits,
  risks and challenges}. In \bibinfo{booktitle}{\emph{Proceedings of the 44th
  ASC Annual Conference}}. \bibinfo{pages}{2--5}.
\newblock


\bibitem[\protect\citeauthoryear{Balaji, Bhattacharya, Fierro, Gao, Gluck,
  Hong, Johansen, Koh, Ploennigs, Agarwal, et~al\mbox{.}}{Balaji
  et~al\mbox{.}}{2016}]%
        {balaji2016brick}
\bibfield{author}{\bibinfo{person}{Bharathan Balaji}, \bibinfo{person}{Arka
  Bhattacharya}, \bibinfo{person}{Gabriel Fierro}, \bibinfo{person}{Jingkun
  Gao}, \bibinfo{person}{Joshua Gluck}, \bibinfo{person}{Dezhi Hong},
  \bibinfo{person}{Aslak Johansen}, \bibinfo{person}{Jason Koh},
  \bibinfo{person}{Joern Ploennigs}, \bibinfo{person}{Yuvraj Agarwal},
  {et~al\mbox{.}}} \bibinfo{year}{2016}\natexlab{}.
\newblock \showarticletitle{Brick: Towards a unified metadata schema for
  buildings}. In \bibinfo{booktitle}{\emph{Proceedings of the 3rd ACM
  International Conference on Systems for Energy-Efficient Built
  Environments}}. \bibinfo{pages}{41--50}.
\newblock


\bibitem[\protect\citeauthoryear{Balaji, Bhattacharya, Fierro, Gao, Gluck,
  Hong, Johansen, Koh, Ploennigs, Agarwal, et~al\mbox{.}}{Balaji
  et~al\mbox{.}}{2018}]%
        {balaji2018brick}
\bibfield{author}{\bibinfo{person}{Bharathan Balaji}, \bibinfo{person}{Arka
  Bhattacharya}, \bibinfo{person}{Gabriel Fierro}, \bibinfo{person}{Jingkun
  Gao}, \bibinfo{person}{Joshua Gluck}, \bibinfo{person}{Dezhi Hong},
  \bibinfo{person}{Aslak Johansen}, \bibinfo{person}{Jason Koh},
  \bibinfo{person}{Joern Ploennigs}, \bibinfo{person}{Yuvraj Agarwal},
  {et~al\mbox{.}}} \bibinfo{year}{2018}\natexlab{}.
\newblock \showarticletitle{Brick: Metadata schema for portable smart building
  applications}.
\newblock \bibinfo{journal}{\emph{Applied energy}}  \bibinfo{volume}{226}
  (\bibinfo{year}{2018}), \bibinfo{pages}{1273--1292}.
\newblock


\bibitem[\protect\citeauthoryear{Becerik-Gerber, Jazizadeh, Li, and
  Calis}{Becerik-Gerber et~al\mbox{.}}{2012}]%
        {becerik2012application}
\bibfield{author}{\bibinfo{person}{Burcin Becerik-Gerber},
  \bibinfo{person}{Farrokh Jazizadeh}, \bibinfo{person}{Nan Li}, {and}
  \bibinfo{person}{Gulben Calis}.} \bibinfo{year}{2012}\natexlab{}.
\newblock \showarticletitle{Application areas and data requirements for
  BIM-enabled facilities management}.
\newblock \bibinfo{journal}{\emph{Journal of construction engineering and
  management}} \bibinfo{volume}{138}, \bibinfo{number}{3}
  (\bibinfo{year}{2012}), \bibinfo{pages}{431--442}.
\newblock


\bibitem[\protect\citeauthoryear{Bhattacharya, Ploennigs, and
  Culler}{Bhattacharya et~al\mbox{.}}{2015}]%
        {bhattacharya2015short}
\bibfield{author}{\bibinfo{person}{Arka Bhattacharya}, \bibinfo{person}{Joern
  Ploennigs}, {and} \bibinfo{person}{David Culler}.}
  \bibinfo{year}{2015}\natexlab{}.
\newblock \showarticletitle{Short paper: Analyzing metadata schemas for
  buildings: The good, the bad, and the ugly}. In
  \bibinfo{booktitle}{\emph{Proceedings of the 2nd ACM International Conference
  on Embedded Systems for Energy-Efficient Built Environments}}.
  \bibinfo{pages}{33--34}.
\newblock


\bibitem[\protect\citeauthoryear{Bostock}{Bostock}{2012}]%
        {d3js}
\bibfield{author}{\bibinfo{person}{Mike Bostock}.}
  \bibinfo{year}{2012}\natexlab{}.
\newblock \bibinfo{title}{D3.js - Data-Driven Documents}.
\newblock
\newblock
\urldef\tempurl%
\url{http://d3js.org/}
\showURL{%
\tempurl}


\bibitem[\protect\citeauthoryear{Bryde, Broquetas, and Volm}{Bryde
  et~al\mbox{.}}{2013}]%
        {bryde2013project}
\bibfield{author}{\bibinfo{person}{David Bryde}, \bibinfo{person}{Mart{\'\i}
  Broquetas}, {and} \bibinfo{person}{J{\"u}rgen~Marc Volm}.}
  \bibinfo{year}{2013}\natexlab{}.
\newblock \showarticletitle{The project benefits of building information
  modelling (BIM)}.
\newblock \bibinfo{journal}{\emph{International journal of project management}}
  \bibinfo{volume}{31}, \bibinfo{number}{7} (\bibinfo{year}{2013}),
  \bibinfo{pages}{971--980}.
\newblock


\bibitem[\protect\citeauthoryear{Cantarero, Rubio, Trapero, Santofimia,
  Villanueva, Villa, and Lopez}{Cantarero et~al\mbox{.}}{2018}]%
        {cantarero2018common}
\bibfield{author}{\bibinfo{person}{Ruben Cantarero}, \bibinfo{person}{Ana
  Rubio}, \bibinfo{person}{Cristian Trapero}, \bibinfo{person}{Maria~J
  Santofimia}, \bibinfo{person}{Felix~J Villanueva}, \bibinfo{person}{David
  Villa}, {and} \bibinfo{person}{Juan~C Lopez}.}
  \bibinfo{year}{2018}\natexlab{}.
\newblock \showarticletitle{A common-sense based system for Geo-IoT}.
\newblock \bibinfo{journal}{\emph{Procedia Computer Science}}
  \bibinfo{volume}{126} (\bibinfo{year}{2018}), \bibinfo{pages}{665--674}.
\newblock


\bibitem[\protect\citeauthoryear{Cascone, Ferrara, Giovannini, and
  Serale}{Cascone et~al\mbox{.}}{2017}]%
        {cascone2017ethical}
\bibfield{author}{\bibinfo{person}{Ylenia Cascone}, \bibinfo{person}{Maria
  Ferrara}, \bibinfo{person}{Luigi Giovannini}, {and} \bibinfo{person}{Gianluca
  Serale}.} \bibinfo{year}{2017}\natexlab{}.
\newblock \showarticletitle{Ethical issues of monitoring sensor networks for
  energy efficiency in smart buildings: a case study}.
\newblock \bibinfo{journal}{\emph{Energy Procedia}}  \bibinfo{volume}{134}
  (\bibinfo{year}{2017}), \bibinfo{pages}{337--345}.
\newblock


\bibitem[\protect\citeauthoryear{Chevallier, Finance, and Boulakia}{Chevallier
  et~al\mbox{.}}{[n.d.]}]%
        {chevallierreference}
\bibfield{author}{\bibinfo{person}{Zo{\'e} Chevallier},
  \bibinfo{person}{B{\'e}atrice Finance}, {and} \bibinfo{person}{Benjamin~Cohen
  Boulakia}.} \bibinfo{year}{[n.d.]}\natexlab{}.
\newblock \showarticletitle{A Reference Architecture for Smart Building Digital
  Twin}.
\newblock  (\bibinfo{year}{[n.\,d.]}).
\newblock


\bibitem[\protect\citeauthoryear{Clement~Escoffier}{Clement~Escoffier}{2015}]%
        {clement2015vert}
\bibfield{author}{\bibinfo{person}{MK Clement~Escoffier}.}
  \bibinfo{year}{2015}\natexlab{}.
\newblock \showarticletitle{Vert. x--a toolkit for building reactive
  applications on the JVM}.
\newblock \bibinfo{journal}{\emph{Online. http://vertx. io}}
  (\bibinfo{year}{2015}).
\newblock


\bibitem[\protect\citeauthoryear{Dave, Buda, Nurminen, and Fr{\"a}mling}{Dave
  et~al\mbox{.}}{2018}]%
        {dave2018framework}
\bibfield{author}{\bibinfo{person}{Bhargav Dave}, \bibinfo{person}{Andrea
  Buda}, \bibinfo{person}{Antti Nurminen}, {and} \bibinfo{person}{Kary
  Fr{\"a}mling}.} \bibinfo{year}{2018}\natexlab{}.
\newblock \showarticletitle{A framework for integrating BIM and IoT through
  open standards}.
\newblock \bibinfo{journal}{\emph{Automation in Construction}}
  \bibinfo{volume}{95} (\bibinfo{year}{2018}), \bibinfo{pages}{35--45}.
\newblock


\bibitem[\protect\citeauthoryear{Doan, Ghaffarianhoseini, Naismith, Zhang,
  Tookey, et~al\mbox{.}}{Doan et~al\mbox{.}}{2019}]%
        {doan2019bim}
\bibfield{author}{\bibinfo{person}{D Doan}, \bibinfo{person}{Ali
  Ghaffarianhoseini}, \bibinfo{person}{Nicola Naismith},
  \bibinfo{person}{Tongrui Zhang}, \bibinfo{person}{T Tookey}, {et~al\mbox{.}}}
  \bibinfo{year}{2019}\natexlab{}.
\newblock \showarticletitle{What is BIM?: a need for a unique BIM definition}.
\newblock  (\bibinfo{year}{2019}).
\newblock


\bibitem[\protect\citeauthoryear{Eastman et~al\mbox{.}}{Eastman
  et~al\mbox{.}}{1974}]%
        {eastman1974outline}
\bibfield{author}{\bibinfo{person}{Charles Eastman} {et~al\mbox{.}}}
  \bibinfo{year}{1974}\natexlab{}.
\newblock \showarticletitle{An Outline of the Building Description System.
  Research Report No. 50.}
\newblock  (\bibinfo{year}{1974}).
\newblock


\bibitem[\protect\citeauthoryear{Fan, Levine, Wen, and Qiu}{Fan
  et~al\mbox{.}}{2017}]%
        {fan2017deep}
\bibfield{author}{\bibinfo{person}{Yaxiang Fan}, \bibinfo{person}{Martin~D
  Levine}, \bibinfo{person}{Gongjian Wen}, {and} \bibinfo{person}{Shaohua
  Qiu}.} \bibinfo{year}{2017}\natexlab{}.
\newblock \showarticletitle{A deep neural network for real-time detection of
  falling humans in naturally occurring scenes}.
\newblock \bibinfo{journal}{\emph{Neurocomputing}}  \bibinfo{volume}{260}
  (\bibinfo{year}{2017}), \bibinfo{pages}{43--58}.
\newblock


\bibitem[\protect\citeauthoryear{Franco and Leccese}{Franco and
  Leccese}{2020}]%
        {franco2020measurement}
\bibfield{author}{\bibinfo{person}{Alessandro Franco} {and}
  \bibinfo{person}{Francesco Leccese}.} \bibinfo{year}{2020}\natexlab{}.
\newblock \showarticletitle{Measurement of CO2 concentration for occupancy
  estimation in educational buildings with energy efficiency purposes}.
\newblock \bibinfo{journal}{\emph{Journal of Building Engineering}}
  (\bibinfo{year}{2020}), \bibinfo{pages}{101714}.
\newblock


\bibitem[\protect\citeauthoryear{Gao, Wang, Xu, Song, and Jia}{Gao
  et~al\mbox{.}}{2019}]%
        {gao2019real}
\bibfield{author}{\bibinfo{person}{Song Gao}, \bibinfo{person}{Hongyu Wang},
  \bibinfo{person}{Fang Xu}, \bibinfo{person}{Jilai Song}, {and}
  \bibinfo{person}{Kai Jia}.} \bibinfo{year}{2019}\natexlab{}.
\newblock \showarticletitle{Real-time Human Action Detection for Elderly
  Monitoring System}. In \bibinfo{booktitle}{\emph{2019 IEEE 9th Annual
  International Conference on CYBER Technology in Automation, Control, and
  Intelligent Systems (CYBER)}}. IEEE, \bibinfo{pages}{1191--1196}.
\newblock


\bibitem[\protect\citeauthoryear{Hassner, Itcher, and Kliper-Gross}{Hassner
  et~al\mbox{.}}{2012}]%
        {hassner2012violent}
\bibfield{author}{\bibinfo{person}{Tal Hassner}, \bibinfo{person}{Yossi
  Itcher}, {and} \bibinfo{person}{Orit Kliper-Gross}.}
  \bibinfo{year}{2012}\natexlab{}.
\newblock \showarticletitle{Violent flows: Real-time detection of violent crowd
  behavior}. In \bibinfo{booktitle}{\emph{2012 IEEE Computer Society Conference
  on Computer Vision and Pattern Recognition Workshops}}. IEEE,
  \bibinfo{pages}{1--6}.
\newblock


\bibitem[\protect\citeauthoryear{Hern{\'a}ndez-Ramos, Moreno, Bernab{\'e},
  Carrillo, and Skarmeta}{Hern{\'a}ndez-Ramos et~al\mbox{.}}{2015}]%
        {hernandez2015safir}
\bibfield{author}{\bibinfo{person}{Jos{\'e}~L Hern{\'a}ndez-Ramos},
  \bibinfo{person}{M~Victoria Moreno}, \bibinfo{person}{Jorge~Bernal
  Bernab{\'e}}, \bibinfo{person}{Dan~Garc{\'\i}a Carrillo}, {and}
  \bibinfo{person}{Antonio~F Skarmeta}.} \bibinfo{year}{2015}\natexlab{}.
\newblock \showarticletitle{SAFIR: Secure access framework for IoT-enabled
  services on smart buildings}.
\newblock \bibinfo{journal}{\emph{J. Comput. System Sci.}}
  \bibinfo{volume}{81}, \bibinfo{number}{8} (\bibinfo{year}{2015}),
  \bibinfo{pages}{1452--1463}.
\newblock


\bibitem[\protect\citeauthoryear{Holten, Rasmussen, Hviid, and
  Karlsh{\o}}{Holten et~al\mbox{.}}{2018}]%
        {holten2018integrating}
\bibfield{author}{\bibinfo{person}{Mads Holten}, \bibinfo{person}{Christian
  Aaskov~Frausing Rasmussen}, \bibinfo{person}{Christian~Anker Hviid}, {and}
  \bibinfo{person}{Jan Karlsh{\o}}.} \bibinfo{year}{2018}\natexlab{}.
\newblock \showarticletitle{Integrating Building Information Modeling and
  Sensor Observations using Semantic Web}.
\newblock  (\bibinfo{year}{2018}).
\newblock


\bibitem[\protect\citeauthoryear{Hui, Sherratt, and S{\'a}nchez}{Hui
  et~al\mbox{.}}{2017}]%
        {hui2017major}
\bibfield{author}{\bibinfo{person}{Terence~KL Hui}, \bibinfo{person}{R~Simon
  Sherratt}, {and} \bibinfo{person}{Daniel~D{\'\i}az S{\'a}nchez}.}
  \bibinfo{year}{2017}\natexlab{}.
\newblock \showarticletitle{Major requirements for building Smart Homes in
  Smart Cities based on Internet of Things technologies}.
\newblock \bibinfo{journal}{\emph{Future Generation Computer Systems}}
  \bibinfo{volume}{76} (\bibinfo{year}{2017}), \bibinfo{pages}{358--369}.
\newblock


\bibitem[\protect\citeauthoryear{Jourdan, Meyer, and Bacher}{Jourdan
  et~al\mbox{.}}{2019}]%
        {jourdan2019towards}
\bibfield{author}{\bibinfo{person}{Matthieu Jourdan}, \bibinfo{person}{Florian
  Meyer}, {and} \bibinfo{person}{Jean-Philippe Bacher}.}
  \bibinfo{year}{2019}\natexlab{}.
\newblock \showarticletitle{Towards an integrated approach of building-data
  management through the convergence of Building Information Modelling and
  Internet of Things}. In \bibinfo{booktitle}{\emph{Journal of Physics:
  Conference Series}}, Vol.~\bibinfo{volume}{1343}. IOP Publishing,
  \bibinfo{pages}{012135}.
\newblock


\bibitem[\protect\citeauthoryear{Khalid, Bashir, and Newport}{Khalid
  et~al\mbox{.}}{2017}]%
        {inbook}
\bibfield{author}{\bibinfo{person}{Muhammad Khalid}, \bibinfo{person}{Muhammad
  Bashir}, {and} \bibinfo{person}{Darryl Newport}.}
  \bibinfo{year}{2017}\natexlab{}.
\newblock \bibinfo{booktitle}{\emph{Development of a Building Information
  Modelling (BIM)-Based Real-Time Data Integration System Using a Building
  Management System (BMS)}}.
\newblock \bibinfo{pages}{93--104}.
\newblock
\showISBNx{978-3-319-50345-5}
\urldef\tempurl%
\url{https://doi.org/10.1007/978-3-319-50346-2_7}
\showDOI{\tempurl}


\bibitem[\protect\citeauthoryear{Koh, Balaji, Sengupta, McAuley, Gupta, and
  Agarwal}{Koh et~al\mbox{.}}{2018}]%
        {koh2018scrabble}
\bibfield{author}{\bibinfo{person}{Jason Koh}, \bibinfo{person}{Bharathan
  Balaji}, \bibinfo{person}{Dhiman Sengupta}, \bibinfo{person}{Julian McAuley},
  \bibinfo{person}{Rajesh Gupta}, {and} \bibinfo{person}{Yuvraj Agarwal}.}
  \bibinfo{year}{2018}\natexlab{}.
\newblock \showarticletitle{Scrabble: transferrable semi-automated semantic
  metadata normalization using intermediate representation}. In
  \bibinfo{booktitle}{\emph{Proceedings of the 5th Conference on Systems for
  Built Environments}}. \bibinfo{pages}{11--20}.
\newblock


\bibitem[\protect\citeauthoryear{Manic, Wijayasekara, Amarasinghe, and
  Rodriguez-Andina}{Manic et~al\mbox{.}}{2016}]%
        {manic2016building}
\bibfield{author}{\bibinfo{person}{Milos Manic}, \bibinfo{person}{Dumidu
  Wijayasekara}, \bibinfo{person}{Kasun Amarasinghe}, {and}
  \bibinfo{person}{Juan~J Rodriguez-Andina}.} \bibinfo{year}{2016}\natexlab{}.
\newblock \showarticletitle{Building energy management systems: The age of
  intelligent and adaptive buildings}.
\newblock \bibinfo{journal}{\emph{IEEE Industrial Electronics Magazine}}
  \bibinfo{volume}{10}, \bibinfo{number}{1} (\bibinfo{year}{2016}),
  \bibinfo{pages}{25--39}.
\newblock


\bibitem[\protect\citeauthoryear{Mousa, Luo, and McCabe}{Mousa
  et~al\mbox{.}}{2016}]%
        {mousa2016utilizing}
\bibfield{author}{\bibinfo{person}{Michael Mousa}, \bibinfo{person}{Xiaowei
  Luo}, {and} \bibinfo{person}{Brenda McCabe}.}
  \bibinfo{year}{2016}\natexlab{}.
\newblock \showarticletitle{Utilizing BIM and carbon estimating methods for
  meaningful data representation}.
\newblock \bibinfo{journal}{\emph{Procedia Engineering}}  \bibinfo{volume}{145}
  (\bibinfo{year}{2016}), \bibinfo{pages}{1242--1249}.
\newblock


\bibitem[\protect\citeauthoryear{Muralidhara and Hegde}{Muralidhara and
  Hegde}{[n.d.]}]%
        {muralidharaair}
\bibfield{author}{\bibinfo{person}{Shishir Muralidhara} {and}
  \bibinfo{person}{Niharika Hegde}.} \bibinfo{year}{[n.d.]}\natexlab{}.
\newblock \showarticletitle{Air Quality Monitoring and Gas Leakage Detection
  with Automatic Shut-Off using Wireless Sensor-Actuator Networks}.
\newblock \bibinfo{journal}{\emph{Internet Technology Letters}}
  (\bibinfo{year}{[n.\,d.]}), \bibinfo{pages}{e185}.
\newblock


\bibitem[\protect\citeauthoryear{NBS}{NBS}{2019}]%
        {nbs2019national}
\bibfield{author}{\bibinfo{person}{NBS}.} \bibinfo{year}{2019}\natexlab{}.
\newblock \bibinfo{title}{National BIM report 2019--the definitive industry
  update.}
\newblock
\newblock


\bibitem[\protect\citeauthoryear{Niu, Long, Han, and Wang}{Niu
  et~al\mbox{.}}{2004}]%
        {niu2004human}
\bibfield{author}{\bibinfo{person}{Wei Niu}, \bibinfo{person}{Jiao Long},
  \bibinfo{person}{Dan Han}, {and} \bibinfo{person}{Yuan-Fang Wang}.}
  \bibinfo{year}{2004}\natexlab{}.
\newblock \showarticletitle{Human activity detection and recognition for video
  surveillance}. In \bibinfo{booktitle}{\emph{2004 IEEE International
  Conference on Multimedia and Expo (ICME)(IEEE Cat. No. 04TH8763)}},
  Vol.~\bibinfo{volume}{1}. IEEE, \bibinfo{pages}{719--722}.
\newblock


\bibitem[\protect\citeauthoryear{{OpenStreetMap contributors}}{{OpenStreetMap
  contributors}}{2017}]%
        {OpenStreetMap}
\bibfield{author}{\bibinfo{person}{{OpenStreetMap contributors}}.}
  \bibinfo{year}{2017}\natexlab{}.
\newblock \bibinfo{title}{{Planet dump retrieved from https://planet.osm.org
  }}.
\newblock \bibinfo{howpublished}{\url{ https://www.openstreetmap.org }}.
\newblock


\bibitem[\protect\citeauthoryear{Park, Kim, and Cho}{Park
  et~al\mbox{.}}{2017}]%
        {park2017framework}
\bibfield{author}{\bibinfo{person}{JeeWoong Park}, \bibinfo{person}{Kyungki
  Kim}, {and} \bibinfo{person}{Yong~K Cho}.} \bibinfo{year}{2017}\natexlab{}.
\newblock \showarticletitle{Framework of automated construction-safety
  monitoring using cloud-enabled BIM and BLE mobile tracking sensors}.
\newblock \bibinfo{journal}{\emph{Journal of Construction Engineering and
  Management}} \bibinfo{volume}{143}, \bibinfo{number}{2}
  (\bibinfo{year}{2017}), \bibinfo{pages}{05016019}.
\newblock


\bibitem[\protect\citeauthoryear{Penna, Regis, Schweigkofler, Marcher, and
  Matt}{Penna et~al\mbox{.}}{2019}]%
        {penna2019sensors}
\bibfield{author}{\bibinfo{person}{Paola Penna}, \bibinfo{person}{Gian~Luca
  Regis}, \bibinfo{person}{Alice Schweigkofler}, \bibinfo{person}{Carmen
  Marcher}, {and} \bibinfo{person}{Dominik Matt}.}
  \bibinfo{year}{2019}\natexlab{}.
\newblock \showarticletitle{From Sensors to BIM: Monitoring Comfort Conditions
  of Social Housing with the KlimaKit Model}. In
  \bibinfo{booktitle}{\emph{International Conference on Cooperative Design,
  Visualization and Engineering}}. Springer, \bibinfo{pages}{108--115}.
\newblock


\bibitem[\protect\citeauthoryear{Perera and Skeie}{Perera and Skeie}{2016}]%
        {perera2016estimation}
\bibfield{author}{\bibinfo{person}{Degurunnehalage Wathsala~Upamali Perera}
  {and} \bibinfo{person}{Nils-Olav Skeie}.} \bibinfo{year}{2016}\natexlab{}.
\newblock \showarticletitle{Estimation of the heating time of small-scale
  buildings using dynamic models}.
\newblock \bibinfo{journal}{\emph{Buildings}} \bibinfo{volume}{6},
  \bibinfo{number}{1} (\bibinfo{year}{2016}), \bibinfo{pages}{10}.
\newblock


\bibitem[\protect\citeauthoryear{Ramprasad, McArthur, Fokaefs, Barna, Damm, and
  Litoiu}{Ramprasad et~al\mbox{.}}{2018}]%
        {ramprasad2018leveraging}
\bibfield{author}{\bibinfo{person}{Brian Ramprasad}, \bibinfo{person}{Jenn
  McArthur}, \bibinfo{person}{Marios Fokaefs}, \bibinfo{person}{Cornel Barna},
  \bibinfo{person}{Mark Damm}, {and} \bibinfo{person}{Marin Litoiu}.}
  \bibinfo{year}{2018}\natexlab{}.
\newblock \showarticletitle{Leveraging existing sensor networks as IoT devices
  for smart buildings}. In \bibinfo{booktitle}{\emph{2018 IEEE 4th World Forum
  on Internet of Things (WF-IoT)}}. IEEE, \bibinfo{pages}{452--457}.
\newblock


\bibitem[\protect\citeauthoryear{Rasmussen}{Rasmussen}{2018}]%
        {rasmussen2018madsholten}
\bibfield{author}{\bibinfo{person}{Mads~Holten Rasmussen}.}
  \bibinfo{year}{2018}\natexlab{}.
\newblock \bibinfo{title}{MadsHolten/revit-bot-exporter export from autodesk
  revit to bot-compliant turtle}.
\newblock
\newblock


\bibitem[\protect\citeauthoryear{Rasmussen, Hviid, and Karlsh{\o}j}{Rasmussen
  et~al\mbox{.}}{2017}]%
        {rasmussen2017web}
\bibfield{author}{\bibinfo{person}{Mads~Holten Rasmussen},
  \bibinfo{person}{Christian~Anker Hviid}, {and} \bibinfo{person}{Jan
  Karlsh{\o}j}.} \bibinfo{year}{2017}\natexlab{}.
\newblock \showarticletitle{Web-based topology queries on a BIM model}. In
  \bibinfo{booktitle}{\emph{LDAC2017--5th Linked Data in Architecture and
  Construction Workshop}}.
\newblock


\bibitem[\protect\citeauthoryear{SmartCambridge}{SmartCambridge}{[n.d.]}]%
        {smartcambridge}
\bibfield{author}{\bibinfo{person}{SmartCambridge}.}
  \bibinfo{year}{[n.d.]}\natexlab{}.
\newblock \bibinfo{title}{SmartCambridge}.
\newblock
\newblock
\urldef\tempurl%
\url{https://smartcambridge.org/}
\showURL{%
\tempurl}


\bibitem[\protect\citeauthoryear{Solutions}{Solutions}{2020}]%
        {sip_platform}
\bibfield{author}{\bibinfo{person}{Synapsys Solutions}.}
  \bibinfo{year}{2020}\natexlab{}.
\newblock \bibinfo{booktitle}{\emph{SIP+ Platform}}.
\newblock
\urldef\tempurl%
\url{https://www.synapsys-solutions.com/products/sip-platform/}
\showURL{%
\tempurl}


\bibitem[\protect\citeauthoryear{Tang, Shelden, Eastman, Pishdad-Bozorgi, and
  Gao}{Tang et~al\mbox{.}}{2019}]%
        {tang2019review}
\bibfield{author}{\bibinfo{person}{Shu Tang}, \bibinfo{person}{Dennis~R
  Shelden}, \bibinfo{person}{Charles~M Eastman}, \bibinfo{person}{Pardis
  Pishdad-Bozorgi}, {and} \bibinfo{person}{Xinghua Gao}.}
  \bibinfo{year}{2019}\natexlab{}.
\newblock \showarticletitle{A review of building information modeling (BIM) and
  the internet of things (IoT) devices integration: Present status and future
  trends}.
\newblock \bibinfo{journal}{\emph{Automation in Construction}}
  \bibinfo{volume}{101} (\bibinfo{year}{2019}), \bibinfo{pages}{127--139}.
\newblock


\bibitem[\protect\citeauthoryear{Tapia, Intille, and Larson}{Tapia
  et~al\mbox{.}}{2004}]%
        {tapia2004activity}
\bibfield{author}{\bibinfo{person}{Emmanuel~Munguia Tapia},
  \bibinfo{person}{Stephen~S Intille}, {and} \bibinfo{person}{Kent Larson}.}
  \bibinfo{year}{2004}\natexlab{}.
\newblock \showarticletitle{Activity recognition in the home using simple and
  ubiquitous sensors}. In \bibinfo{booktitle}{\emph{International conference on
  pervasive computing}}. Springer, \bibinfo{pages}{158--175}.
\newblock


\bibitem[\protect\citeauthoryear{Teizer, Wolf, Golovina, Perschewski, Propach,
  Neges, and K{\"o}nig}{Teizer et~al\mbox{.}}{2017}]%
        {teizer2017internet}
\bibfield{author}{\bibinfo{person}{Jochen Teizer}, \bibinfo{person}{Mario
  Wolf}, \bibinfo{person}{Olga Golovina}, \bibinfo{person}{Manuel Perschewski},
  \bibinfo{person}{Markus Propach}, \bibinfo{person}{Matthias Neges}, {and}
  \bibinfo{person}{Markus K{\"o}nig}.} \bibinfo{year}{2017}\natexlab{}.
\newblock \showarticletitle{Internet of Things (IoT) for integrating
  environmental and localization data in Building Information Modeling (BIM)}.
  In \bibinfo{booktitle}{\emph{ISARC. Proceedings of the International
  Symposium on Automation and Robotics in Construction}},
  Vol.~\bibinfo{volume}{34}. IAARC Publications.
\newblock


\bibitem[\protect\citeauthoryear{Teo and Yu}{Teo and Yu}{2017}]%
        {teo2017extraction}
\bibfield{author}{\bibinfo{person}{Tee-Ann Teo} {and} \bibinfo{person}{Sz-Cheng
  Yu}.} \bibinfo{year}{2017}\natexlab{}.
\newblock \showarticletitle{THE EXTRACTION OF INDOOR BUILDING INFORMATION FROM
  BIM TO OGC INDOORGML.}
\newblock \bibinfo{journal}{\emph{International Archives of the Photogrammetry,
  Remote Sensing \& Spatial Information Sciences}}  \bibinfo{volume}{42}
  (\bibinfo{year}{2017}).
\newblock


\bibitem[\protect\citeauthoryear{{three.js}}{{three.js}}{2015}]%
        {noauthororeditor2015threejs}
\bibfield{author}{\bibinfo{person}{{three.js}}.}
  \bibinfo{year}{2015}\natexlab{}.
\newblock \bibinfo{title}{three.js / editor}.
\newblock
\newblock
\urldef\tempurl%
\url{http://threejs.org/editor/}
\showURL{%
\tempurl}


\bibitem[\protect\citeauthoryear{Voigt and Von~dem Bussche}{Voigt and Von~dem
  Bussche}{2017}]%
        {voigt2017eu}
\bibfield{author}{\bibinfo{person}{Paul Voigt} {and} \bibinfo{person}{Axel
  Von~dem Bussche}.} \bibinfo{year}{2017}\natexlab{}.
\newblock \showarticletitle{The eu general data protection regulation (gdpr)}.
\newblock \bibinfo{journal}{\emph{A Practical Guide, 1st Ed., Cham: Springer
  International Publishing}} (\bibinfo{year}{2017}).
\newblock


\bibitem[\protect\citeauthoryear{Zhang, Kuppannagari, Xiong, Kannan, and
  Prasanna}{Zhang et~al\mbox{.}}{2019}]%
        {zhang2019cooperative}
\bibfield{author}{\bibinfo{person}{Chi Zhang}, \bibinfo{person}{Sanmukh~R
  Kuppannagari}, \bibinfo{person}{Chuanxiu Xiong}, \bibinfo{person}{Rajgopal
  Kannan}, {and} \bibinfo{person}{Viktor~K Prasanna}.}
  \bibinfo{year}{2019}\natexlab{}.
\newblock \showarticletitle{A cooperative multi-agent deep reinforcement
  learning framework for real-time residential load scheduling}. In
  \bibinfo{booktitle}{\emph{Proceedings of the International Conference on
  Internet of Things Design and Implementation}}. \bibinfo{pages}{59--69}.
\newblock


\bibitem[\protect\citeauthoryear{Zhang and Ardakanian}{Zhang and
  Ardakanian}{2019}]%
        {zhang2019domain}
\bibfield{author}{\bibinfo{person}{Tianyu Zhang} {and} \bibinfo{person}{Omid
  Ardakanian}.} \bibinfo{year}{2019}\natexlab{}.
\newblock \showarticletitle{A domain adaptation technique for fine-grained
  occupancy estimation in commercial buildings}. In
  \bibinfo{booktitle}{\emph{Proceedings of the International Conference on
  Internet of Things Design and Implementation}}. \bibinfo{pages}{148--159}.
\newblock


\bibitem[\protect\citeauthoryear{Zucker, Ferhatbegovic, and Bruckner}{Zucker
  et~al\mbox{.}}{2012}]%
        {zucker2012building}
\bibfield{author}{\bibinfo{person}{Gerhard Zucker}, \bibinfo{person}{Tarik
  Ferhatbegovic}, {and} \bibinfo{person}{Dietmar Bruckner}.}
  \bibinfo{year}{2012}\natexlab{}.
\newblock \showarticletitle{Building automation for increased energy efficiency
  in buildings}. In \bibinfo{booktitle}{\emph{2012 IEEE International Symposium
  on Industrial Electronics}}. IEEE, \bibinfo{pages}{1191--1196}.
\newblock


\end{thebibliography}
}

\end{document}